# A comprehensive spectroscopic reference of the solar system and its application to exoplanet direct imaging

**Allison Payne[1,2], Geronimo L. Villanueva[1,*], Vincent Kofman[3,1,4,], Thomas J. Fauchez[1,4], Sara Faggi[1,4], Avi M. Mandell[1], Aki Roberge[1], Eleonora Alei[1,5]**

1) NASA Goddard Space Flight Center, 8800 Goddard Rd, Greenbelt MD, USA

2) Southeastern Universities Research Association (SURA), Greenbelt MD, USA

3) Centre for Planetary Habitability, Department of Geosciences, University of Oslo, Oslo, Norway.

4) American University, Washington DC, USA

5) Oak Ridge Associated Universities (ORAU), NPP Program, Oak Ridge TN, USA

* Corresponding author: Geronimo L. Villanueva (geronimo.villanueva@nasa.gov)

## Abstract

We present a calibrated database of reflectance spectra for the solar system planets (i.e., Mercury, Venus, Earth, Mars, Jupiter, Saturn, Uranus, Neptune) and for Titan, spanning from the ultraviolet to the near infrared. We considered data collected over 60 years of planetary observations, employing a broad range of geometries and facilities (spacecraft and ground-based observatories). To correct for differences in observational geometries and data quality, we adopted a two-step calibration process that standardized each spectrum to the planet's geometric albedo values and corrected for planetary heterogeneity and calibration effects. The calibrated datasets were then combined across wavelengths, leading to a reference composite reflectance spectrum for each planet. As an application of this spectral library for exoplanetary research, we simulated direct imaging observations of the Proxima Centauri and HD 219134 systems as solar system analogs, as well as the solar system at a distance of 10 parsecs. We also explored the detection limitations of direct imaging instruments imposed by the inner and outer working angles for Earth and Jupiter-like exoplanets as a function of system distance. Additionally, we used the visible light portion of the results to produce realistic color reconstructions of each planet. Standardizing reflectance spectra in this work improves our baseline for interpreting new reflected light observations of exoplanets through comparative planetology. This spectral library can then serve as a calibrated and validated reference in the modeling and preparation for the characterization of exoplanet atmospheres with future direct imaging missions and for astronomical studies of the solar system.

## 1 Introduction

Planetary spectroscopy is a powerful tool for remotely characterizing the physical and chemical properties of exoplanets and their atmospheres. By analyzing the wavelength-dependent absorption, emission, and scattering of electromagnetic radiation as it interacts with planetary atmospheres, astronomers can infer critical parameters including atmospheric composition, temperature profiles, pressure structures, cloud properties, and even potential biosignatures—signs of life. Space-based observatories (e.g., Hubble, JWST) have detected water vapor, carbon dioxide, methane, and other molecular species in distant exoplanetary atmospheres through transmission and emission spectroscopy techniques (e.g., Deming et al. 2013, Lustig-Yaeger, J., et al. 2023, Ahrer et al. 2023). Looking toward the future, the planned Habitable Worlds Observatory (HWO, **Feinberg et al. 2024**) promises to extend these capabilities dramatically, offering unprecedented sensitivity and spectral resolution to detect Earth-like atmospheric signatures around potentially habitable exoplanets orbiting Sun-like stars. This technological progression from current observatories to next-generation facilities would represent a paradigm shift in our ability to not only catalog exoplanetary diversity but to identify worlds that may harbor the conditions necessary for life, fundamentally advancing both planetary science and astrobiology in the coming decades.

As we search for spectroscopic signatures on distant worlds, having a strong reference spectroscopic framework will be critical. The intent of this work is then to calibrate and combine existing solar system reflectance spectra to establish a ground truth dataset that is applicable in the context of atmospheric modeling and future direct imaging of exoplanets. The study leverages the diversity of planets in our solar system to produce a calibrated spectral library enabling direct comparison across planetary types, upon calibration to geometric albedo values in this work. The framework provided here can be then used to test the reliability of synthetic data and improve our tools for reproducing observations including radiative transfer simulations, thus understanding how planets may appear under different viewing phases, angles, realistic modeling of clouds, surface albedo, and temporal changes such as diurnal or seasonal effects. Additionally, with ongoing efforts to find and characterize Earth-like planets it is essential that we have a strong understanding of how Earth may appear in a range of observational configurations (e.g., Sagan et al. 1993; Livengood et al., 2011; Robinson et al., 2010, 2011; Kofman et al., 2024; Strauss et al. 2024). Models aid us in predicting how each of our solar system planets may be observed from light years away and are crucial in our preparation for upcoming missions that intend to do so.

This study is the result of a wide literature investigation and a large analysis of existing reflectance spectra. Observations have been compiled from sources dating back nearly 60 years to provide the most comprehensive analysis possible. Our solar system has been studied extensively, allowing a unique opportunity to create an advanced culmination of the efforts of the scientific community. Previous efforts have been made (Lundock et al. 2009, Roberge et al. 2017, Madden and Kaltenegger 2018) to create compiled references for the solar system planets across UV, VIS and IR wavelengths and focused on specific datasets and/or planets. These studies are extremely important because they outline the diversity of atmospheres that may be encountered when searching for habitable planets. Additionally, efforts have been made to provide realistic spectra of Earth across geologic timescales (Meadows 2006, Kaltenegger et al. 2007, Sanromá et al. 2014, Arney et al. 2016, Rugheimer and Kaltenegger 2018), with this work intending to provide a calibration reference framework for these models.

This work considers how existing data may be combined, despite the variability in terms of observational configurations and units the reflectance spectra may be presented in. The key challenge that this paper addresses is that the data that are currently available are in most cases not directly comparable. Data in this work were collected from publicly available reflected light spectra of each of our candidates. The range of sources included here were intended to be broad to ensure that we are considering all the available information for each of the candidates. The breadth of this study ultimately allows a wider spectral extent to be included in our analysis, generally ranging from the UV to the NIR (up to 2.5 μm), that specifically aligns well with the design goals for HWO. Since data are considered from many different observations there are inherent differences in the data that must be accounted for before making direct comparison across the observations. The results of this work can be adopted as a baseline to model solar system planets or exoplanets that may be similar in nature to the ones found in our solar system. In particular with upcoming direct imaging missions, it is especially important that we are able to effectively model and calibrate our framework to be applicable in this context. Additionally, having a strong reference spectrum can also improve our understanding of necessary model complexity. Defining the minimum requirements to model planetary characteristics will ultimately improve efficiency in modeling exoplanets, allowing more candidates to be studied or characterized in a shorter time.

The outline of this paper is as follows: Sections 2 describes the methods employed across the different sections, while sections 3 through 11 will focus on Mercury, Venus, Earth, Mars, Jupiter, Saturn, Titan, Uranus, and Neptune respectively. Each of the planetary sections provides a brief overview of the planetary environment, significant missions, a description of the available reflected light spectra across a range of ground and space-based observations, and the steps taken in our analysis to produce a composite reflectance spectrum. For each section, the composite spectrum may be downloaded and adopted as a reference spectrum for similar planetary types. Section 12 adopts the calibrated reflectance spectra from our work and presents a transformation of our solar system results into analog spectra in the context of exoplanetary systems. This section also provides a discussion of the observational limitations and challenges for direct imaging missions. The atmospheres of Earth and Jupiter are simulated as exoplanets observed at quadrature, and we discuss how observations of their atmospheres are intrinsically dependent on both the inner and outer working angles of a coronagraph and the distance to an exoplanetary system. We also compare the absolute fluxes (Jy) of each of the planets in our study with synthetic exoplanet spectra for Proxima Centauri and HD 219134. The composite spectra are also used to construct true color visuals for each of the solar system planets and Titan, as they would be perceived by the human eye and we consider the value of color characterization in exoplanet observations. Finally, the conclusions section discusses and summarizes our results.

## 2. Methods

The spectra of each object are analyzed both morphologically and photometrically (i.e., in terms of absolute fluxes). The morphological characteristics are essential for understanding compositional variations as sampled by different datasets, while the photometric information provides insights not only into composition (e.g., dark versus bright surface materials) but also into geometric and specific absolute calibration effects associated to each dataset. In this work, all datasets are scaled and calibrated to allow direct comparison, adopting the common unit of geometric albedo to express fluxes. Geometric albedo is defined as the ratio between the flux observed from a planet at full phase and the flux that would be observed from a perfectly scattering, flat disk of the same size and at the same distance, viewed face-on.

For datasets originally obtained at nonzero phase angles or not sampling the fully illuminated disk, a geometric correction is applied to adjust the measured intensity to that corresponding to full illumination (phase angle = 0). We use the Lommel–Seeliger scattering model to correct Mercury datasets and the Lambert model for all other objects. These corrections provide only a first-order approximation to the more complex problem of phase correction, since both models describe single-scattering reflectance, which can be limiting for planets with thick, multiple-scattering atmospheres. The Lommel–Seeliger model is specifically chosen for Mercury because it effectively represents scattering from dark, weakly reflective surfaces like Mercury's. The integral phase function of this model is described in Equation 1 and for Lambert in equation 2, where $\alpha$ represents the phase angle (with $\alpha = 0$ corresponding to opposition, when the planet's full-disk is visible).

$$\Phi(\alpha)_{LS} = \mathrm{P}(\alpha)\left[1 - \sin\frac{\alpha}{2}\tan\frac{\alpha}{2}\ln\left(\cot\frac{\alpha}{4}\right)\right] \quad (1)$$

$$\Phi(\alpha)_{L} = \frac{1}{\pi}\left(\sin\alpha + (\pi - \alpha)\cos\alpha\right) \quad (2)$$

These equations are used as first scaling factors for observations taken at other phase angles and normalized to 1 at zero phase. Note that the geometric albedo of a perfectly white Lambertian sphere is 2/3 at zero phase, while that of a Lommel–Seeliger sphere is 1/8.

In addition, some discrepancies between datasets cannot be corrected using a simple scale factor—such as localized reflectance variations or instrumental inconsistencies—which may arise from planetary heterogeneity or calibration differences. To address these, we apply an additional empirical scaling step, adjusting each dataset to match the intensity of a reference photometric spectrum. The reference photometric spectrum for each object is selected based on: (i) its apparent quality when compared with other sources, (ii) the similarity of its observing geometry to our reference geometry (i.e., full-disk illumination at phase zero, representing geometric albedo), and (iii) the robustness of its calibration methodology. Each reference dataset is introduced in the corresponding section describing the data sources for each object, followed by other datasets presented in chronological order. While modeling is employed in parts of this analysis, it is important to emphasize that all reference photometric spectra are derived from real observations. After each dataset is adjusted, we construct a final composite continuum by combining multiple sources. This approach yields the most reliable composite spectrum and enhances the spectral coverage of the reflected-light data for each planet.

## 3 Mercury

### 3.1 Mercury Introduction

Mercury can be described as an "airless" rocky body with an extremely tenuous atmosphere, yet its relatively plain spectrum still plays an important role in future planetary studies. This work is especially significant for objects greatly impacted by the host star, especially in the form of solar winds. Understanding the type of reflectance signatures associated with a Mercury-like environment could aid the process of distinguishing airless rocky planets and characterizing planets seen through future coronagraph missions. It is the smallest planet orbiting the sun, with a diameter of approximately 40% of the Earth and only about 5% of Earth's mass. Mercury has an uncompressed, or zero pressure, density of 5.4 g/cm$^3$ (Anderson et al. 1987) which is larger than both Earth and Venus (4.4 g/cm$^3$ for both) (Nittler et al. 2017). The high density implies a very large, mostly iron core, and the presence of a magnetic field leads us to believe that the core has not yet completely solidified. The magnetic field detection made by Mariner 10 is believed to be a result of the internal dynamo (Balogh and Giampieri, 2002). Mercury particularly stands out from the other terrestrial planets due to its virtually nonexistent atmosphere and surface-based exosphere. These characteristics were first confirmed by the Mariner 10 mission in 1974 (Broadfoot et al. 1976) and the lack of atmospheric insulation leaves Mercury exposed to extreme temperature fluctuations.

Ground-based observations of Mercury are primarily limited to dusk and twilight hours, when the planet is visible, and the sun lies below the horizon. The maximum angular separation between the planet and the Sun is only about 28°, significantly limiting ground-based observations. Given Mercury's relatively low magnitude compared to the Sun and the daytime sky, these observing conditions are necessary to enhance the contrast enough for Mercury to be detected. Sending spacecraft is also very challenging due to the planet's position in the gravity well of the sun and the extremely harsh thermal environment (Stern and Vilas 1988). Despite these obstacles, Mariner 10 became the first spacecraft to explore Mercury, with its closest approach in 1974. This mission confirmed the presence of a weak atmosphere, studied the surface and physical characteristics, and imaged approximately 45% of the planet (Solomon et al. 2001). Mariner 10's greatest scientific contributions were the unprecedented visuals of the planet, revealing its cratered lunar-like surface, and even surprising scientists with the detection of a magnetic field (Dunne & Burgess 1978). Over 30 years passed before the next mission, the MErcury Surface, Space ENvironment, GEochemistry, and Ranging (MESSENGER) probe arrived in 2008 (Solomon et al. 2001, Gold et al. 2001). During its mission the spacecraft collected reflectance data for the planet that has been important in this study. To date this is the only additional spacecraft to reach the planet, and the first to orbit it. The most recent Mercury mission, known as BepiColombo, is being jointly carried out by the European Space Agency (ESA) and the Japanese Aerospace Exploration Agency (JAXA) (Benkhoff et al. 2010). It was launched on October 20th, 2018 and will enter the planet's orbit in late 2025. It aims to conduct a more extensive exploration of the planet and its environment through a study of its interior, surface, exosphere, and magnetosphere (Benkhoff et al. 2021).

### 3.2 Mercury Data

**Mercury Composite Spectrum:** The reflectance spectra in this section are gathered from various sources, as measured employing ground-based observatories (McCord and Clark 1979, Mallama et al. 2017) and from space (MESSENGER; Domingue et al. 2010, Izenberg et al. 2014, Klima et al. 2018), which are then combined to produce a composite spectrum, that may be downloaded as a template for Mercury. The composite spectrum stitches data from Izenberg et al. 2014 [0.311-1.400 μm] and McCord and Clark 1979 [1.410-2.200 μm] to provide a geometric albedo continuum and improve the spectral extent of the individual sources. The data becomes more insightful when it is properly calibrated to extend the spectrum and observe the spectral signature at a wider wavelength range.

**MC1979 Observations:** Mercury integral disc data were collected at the University of Hawaii 2.24 m telescope on Mauna Kea, Hawaii on April 21, 1976 and published in McCord and Clark 1979. These ground-based data cover a spectral range of 0.65 to 2.50 μm, however the observations beyond 2.1 μm were too noisy and excluded in this work. Mercury was observed at a phase angle of 77.4° and the reflectance was reported as an I/F ratio. In this work, the data were adjusted to a phase angle of 0° by applying a geometric scaling factor of 2.102 in Figure 2B. This correction factor was determined using the Lommel-Seeliger surface scattering model. In Figure 2C, the data were multiplied by an empirical scaling factor of 0.66 to align the continuum with the reference geometric albedo data from Mallama et al. 2017 (see below M2017). MC1979 is displayed in Figure 2 and for each of the sources described in this section, the coefficients and observational parameters are recorded in Table 1.

**D2010 Observations:** Observations from Domingue et al. 2010 were collected with the MESSENGER spacecraft, using the onboard Mercury Atmospheric and Surface Composition Spectrometer (MASCS), spanning the ultraviolet-visible range, and the Visible and Infrared Spectrograph (VIRS). MASCS contains both the Ultraviolet and Visible Spectrometer (UVVS) and VIRS. VIRS was created to measure reflectance from the surface of Mercury and provides the data included in this work. The data covers a spectral range of 0.308 to 1.460 μm and was originally published as an I/F ratio. Reflectance values were observed at a phase angle of 74° and corrected to phase zero in Figure 2B by multiplying the data by 1.986, as determined using the Lommel-Seeliger (LS) phase integral. In Figure 2C the data were multiplied by an empirical scaling factor of 4.5 to align the continuum with the reference geometric albedo data from Mallama et al. 2017 (see below M2017).

**I2014 Observations:** Reflectance spectra from Izenberg et al. 2014 were collected using the MASCS instruments onboard the MESSENGER spacecraft. An average reflectance spectrum was determined in their work by considering spectra from locations that represent a wide range of morphologic and color variations. The planetary mean spectrum in their work was derived from more than 850,000 spectra from across the planet. The mean data were published as a unitless corrected reflectance spectrum spanning 0.311 to 1.421 μm. In this work an empirical scale factor is applied to align the data with the reference continuum from Mallama et al. 2017 and in Figure 2C the data were multiplied by 2.

**M2017 Observations (photometric reference):** Geometric albedo values are determined for Mercury within the Johnson Cousins photometric system in Mallama et al. 2017. Mercury was observed with the Solar and Heliospheric Observatory (SOHO) satellite and ground-based instrumentation in Mallama et al. 2002 and the data were reconsidered to calculate the geometric albedo values that were adopted here from Mallama et al. 2017. Data from this source were well calibrated and standardized for each of the planets in the solar system, which is why, for our study, this source was chosen as the reference to which we scaled all the other sources to. The geometric albedo values are displayed in Figure 2 with horizontal bars representing the filter width for each of the values. Data span 0.360 to 0.900 μm across 7 different wavelength bands and are centered at the effective wavelengths for each filter.

**K2018 Observations:** Spectral properties of localized low and high reflectance material (LRM and HRM, respectively) regions are studied over the surface by Klima et al. 2018. Observations were obtained with the MESSENGER spacecraft and the spectra in K2018 originated from the Mercury Dual Imaging System (MDIS) 665 m/pixel final calibrated mosaic of the planet's surface, published in Denevi et al. 2017. Data for this mosaic were collected during the orbital phase of MESSENGER's mission at Mercury between March 18, 2011 and April 30, 2015. In Klima et al. 2018, spectra for various LRM are compared to the HRM region; in this work, the LRM data are adopted from Region 2 in the original publication. The LRM Measurements are from localized regions and do not represent a full-disk view of the planetary surface, so the empirical scale factors adopted here are necessary to align the data with our reference spectrum. In Figure 2C the HRM spectrum is multiplied by a scale factor of 2 to align it with the reference data from M2017, while the LRM is multiplied by a scale factor of 3.2.

## 4 Venus

### 4.1 Venus Introduction

Venus sits about 30% closer to the Sun than Earth, has a similar density, and is the planet in the solar system most similar to our own in terms of size. Venus host's the most massive planetary atmosphere of all the terrestrial planets, composed of approximately 96% carbon dioxide. This large abundance of $CO_2$ contributes to Venus being the hottest planet in the solar system, with a mean temperature of 467° C (~740 K) and surface pressure of about 92 bars. The harsh environment presents significant challenges for reaching the surface and dramatically shortens the lifespan of visiting spacecraft. A key component of Venus's astrobiological relevance lies in its potential for past habitability. It is possible that Venus once accreted an ocean's worth of water, but as it was lost in the unstable environment Venus developed its modern $CO_2$ dominated atmosphere, ruled by a runaway greenhouse effect (Bullock 2013, Ingersoll 1969). This enhances its relevance as an analog for exoplanets with $CO_2$ dominated atmospheres, as well as the importance to learn about how the climate and atmosphere evolved in time, which is of particular interest today due to increasing focus on anthropogenic climate change impacts.

Venus's atmospheric composition has been studied broadly over the past few decades, through ground-based and space-based spectroscopy. Surveys have covered a wide range of wavelengths, and a uniform cloud cover was suggested early on, due to the lack of features at visible wavelengths and the overall brightness (Esposito et al. 1983). Venus's surface was investigated with Earth-based radar observations that allowed instruments to penetrate the atmospheric cloud deck, providing the first maps of the planet's surface, and characterization of mean rotation rate of the planet (Campbell 2019). Despite significant progress in atmospheric characterization of Venus, there are still many unanswered questions and unsolved mysteries regarding its global climate and chemical cycles of its trace gases.

Since the Mariner 2 mission in 1962, Venus has seen 31 successful missions, each providing further insights into the planet's atmosphere and surface conditions. In addition to each of the missions included in this study, Venus will be the target of a series of upcoming missions such as ESA's Venus Orbiter, EnVision (Ghail et al. 2012), and NASA's Deep Atmosphere Venus Investigation of Noble gases, Chemistry, and Imaging (DAVINCI) (Garvin et al. 2022) probe and the Venus Emissivity, Radio Science, InSAR, Topography, And Spectroscopy (VERITAS) (Smrekar et al. 2022) orbiter. Upcoming missions to Venus will enhance our understanding of the planet's evolution and provide information that can aid in understanding if the planet may have been Earth-like previously. Today the planet remains classified as a hot, inhospitable planet, with a dominant $CO_2$ atmosphere, and a global cloud cover of sulfuric acid, whose climatology atmospheric cycles remain not fully understood (O'Rourke et al. 2023).

### 4.2 Venus Data

**Venus Composite Spectrum:** The ground and spaced based observations in this section are combined to produce a composite spectrum for Venus, spanning 0.202 to 1.492 μm. The composite spectrum stitches data from Vlasov et al. 2019 [0.202- 0.309 μm] and Perez-Hoyos [0.311- 1.492 μm]. In our study, the photometric data from Mallama et al. (2017) serve as the key reference spectrum, and we scaled the other sources relative to it. Spectra that overlap with the flux reference point at 0.55 μm are scaled to match this value and data that does not include a flux value at 0.55 μm is scaled by visually aligning the continuum in Figure 3C. The composite spectrum contains scaled versions of the original data, using the factors described below and in Table 2.

**I1968 Observations:** Observations from Irvine 1968 were collected with the Harvard College Observatory's Boyden station in South Africa between 1963 and 1965. Interference filters (with half widths between 5 and 20 nm) were used to focus on 10 narrow bandpasses between 0.315 to 1.06 μm. In their work, data were gathered across a wide range of phase angles and are converted to spherical albedo values. To account for Lambertian limb darkening of the planet, we multiplied the spherical albedo values by 0.66 (Figure 3B). Data in Figure 3C are multiplied by 1.1 to align the continuum with the geometric albedo value at 0.55 μm as presented in Mallama et al. 2017 (see below M2017).

**W1972 Observations:** Observations in Wallace et al. 1972 were collected with the Wisconsin Experiment Package onboard the second Orbiting Astronomical Observatory (OAO-2) (Code et al. 1970), following its launch in December 1968. The albedo values were recorded at a phase angle of 103° and the data are corrected to phase zero in this work. In Figure 3B the data are multiplied by 4.674, using the corresponding phase coefficient for a Lambertian sphere. In Figure 3C, the continuum was multiplied by an empirical scaling factor of 0.85 to best align with the reference spectrum Mallama et al. 2017 (see below M2017) . The author of this publication describes the Venus spectrum as "lower quality data", and it is included here solely to provide a broader range of reference sources.

**B1975 Observations:** Observations from Barker et al. 1975 were collected using the McDonald Observatory's 2.7 m telescope. Integrated disk data covers a spectral range of 0.308- 0.592 μm, by combining data over 6 different observing nights from May to July 1974, producing Venus/ Sun ratio spectra. The spectrum is published in units of spherical albedo and to account for the Lambertian limb darkening on Venus it is multiplied by a scaling factor of 0.66 in Figure 3B. An additional empirical scaling factor of 1.11 is applied in Figure 3C to align the continuum with the geometric albedo value at 0.55 μm as presented in Mallama et al. 2017.

**M1983 Observations:** Data from Moroz 1983 were gathered from various sources and instruments that were combined in their analysis. The author published albedo values as function of wavelengths over a broad range from 0.32 to 3.80 μm, and he applied several approximations and calculations, combining the spectral data and scaling magnitudes to common phase angles of zero (Moroz 1983). For our study, it is important to notice that this source includes data from Barker et al 1975 and Irvine 1968, covering wavelengths ~ 0.38 – 1 μm,, and details on these two sources are provided in the related section B1975 and I1968 respectively. For values reported at wavelengths beyond 1 μm, original albedo data were taken from various sources and combined and synthesized to produce the single scattering albedo values adopted in this work. Details about the original datasets and their analysis are provided in (Moroz et al 1983). In our study we collect the final rescaled values of single scattering albedo as reported in Table 3 of Moroz 1983 and they are shown in Figure 3.

**M2017 Observations (photometric reference):** Geometric albedo values are determined for Venus within the Johnson Cousins photometric system in Mallama et al. 2017. This team initially used a ground based 2000 mm focal length Schmidt–Cassegrain telescope and applied a technique to reduce CCD photometry obtained during daylight hours, however the observable phase angles were limited, and space-based observations were also necessary. Additional observations were obtained using the Large Angle Spectrometric Coronagraph (LASCO) instrument (Brueckner et al., 1995) on the Solar and Heliospheric Observatory (SOHO) (Domingo et al. 1995, Domingo et al. 1994). The author defines the geometric albedo as the ratio of a planet's observed flux to that from a perfectly reflecting Lambert disk of the same size and distance at a phase angle of zero which is slightly different from ours, as we assume a fully illumined flat disk to be the reference. Observations from these sources were well calibrated and standardized for each of the planets in the solar system (Mallama et al. 2017), which is why the visual geometric albedo value of 0.689 at 0.549 μm was chosen as the reference value to which we scaled all the other sources to in this section. The geometric albedo values are displayed in Figure 3 with horizontal bars representing the filter width for each of the values, centered at the effective wavelengths for each Johnson Cousins band.

**PH2018 Observations:** Perez-Hoyos et al. (2018) analyzed data collected during MESSENGER's second Venus flyby on June 5, 2007, on its journey to Mercury. Data were collected by the Visible and InfraRed Spectrograph (VIRS) in the Mercury Atmospheric and Surface Composition Spectrometer (MASCS) (McClintock and Lankton 2007) instrument and a reflected light spectrum is published in Perez-Hoyos et al. 2018. The MASCS instrument includes a Cassegrain telescope with a 257-mm effective focal length and a 50-mm aperture. This supplies information directly to the VIRS instrument, capable of collecting reflectance spectra at spatial scales down to 5 km or less at its closest approach. The MESSENGER observations targeted the planet's equatorial atmosphere, from the subsolar longitude to the terminal, acquiring data across approximately 7500 km along the planet's radius. Their work uses the Non-linear optimal Estimator for MultivariatE spectral analysis (NEMISIS; Irwin et al. 2008) radiative transfer model to provide a best fit of the observed radiance spectra and generate the I/F continuum that is included in this section. Data were initially recorded with a spectral resolution of 4.7 nm and spectral noise was reduced in this work smoothing the continuum by factor of 2. The I/F spectrum covers 0.306-1.492 μm and an empirical scaling factor of 1.13 is applied in Figure 3C to align the MESSENGER data with the accepted visual geometric albedo value of 0.689 for Venus, observed at 0.55 μm (Mallama et al. 2017). The smoothed data were included in the composite spectrum for Venus.

**V2019 Observations:** Initial results from Spectroscopy for Investigation of Characteristics of the Atmosphere of Venus (SPICAV) (Bertaux et al. 2007) onboard the Venus Express (Svedhem et al. 2007) spacecraft are included in Vlasov et al. 2019. The Venus Express spacecraft arrived at the planet on April 11, 2006 and collected data from 2006-2014. SPICAV was designed to study the atmosphere from ground level to the outermost hydrogen corona at 40,000 km (Svedhem et al. 2007) and covers a spectral range of 0.18-0.31 μm. In our study, we only include the nadir data collected on 9 August 2006, covering a spectral range from 0.202-0.309 μm. The data were not scaled in the final plot and agrees well with the intensities described in the related data sets. Data from this source are included in the composite spectrum for Venus.

**L2021 Observations:** Observations in Lee et al. 2021 were collected using JAXA's Akatsuki spacecraft (Venus Climate Orbiter), after a successful orbital insertion on December 7, 2015. Akatsuki was designed to study the planet's atmospheric dynamics and explore cloud physics (Nakamura et al. 2011). The ultraviolet imager (UVI) detected reflected sunlight from the planet with narrow bandpasses focused at 283 nm and 365 nm and a filter width of 14 nm (Yamazaki et al. 2018). The study by Lee et al. 2021 considers full-disk images of the planet taken between December 2015 and January 2019. Disk integrated albedos were produced from nearly 6000 images at each of the wavelengths with phase angles between 3° and 152° and mean phase curves are produced using averaged data over the observation period. In this work we only consider disk integrated albedo values below 10°. All the albedo values for each of the bandpasses were scaled to a phase angle of zero using the Lambert phase integral equation. The average scaled values for each are included as the final geometric albedo and no empirical factors needed to be applied. The geometric albedo value is 0.182 at 283 nm and 0.321 at 365 and are both displayed in Figure 3.

## 5 Earth

### 5.1 Earth Introduction

Earth is the only known planet to sustain life, so it is essential that we develop a reliable reflectance spectrum. The atmospheric signature of Earth will be a key reference for identifying habitable exoplanets in upcoming missions. The atmosphere is composed of approximately 78% $N_2$, 21% $O_2$, less than 1% argon and various trace gases. Earth-like atmospheres are by definition associated with habitability and contain high molecular weight gases that include a condensable greenhouse gas (such as $H_2O$), a non-condensable greenhouse gas (such as $CO_2$), and a non-condensable background gas (such as $N_2$) (Schwieterman et al. 2018). Perhaps surprisingly, few remote observation sources are available that have studied the full-disk spectra of the Earth in reflected light, using methods that would compare directly to how we would observe exoplanets. Obtaining this data requires the instruments to be positioned sufficiently far enough from Earth to view the whole disk of the planet. In this section we consider data from several sources, but we focus on datasets coming from two missions that have observed full-disk observations of the Earth using broad-band spectroscopy. The first study comes from the extended mission of the Deep Impact comet that provides observations of the Earth-Moon system, which were presented in Livengood et al. (2011). The mission obtained photometric data of Earth's partially illuminated disk while observing the planet for 24 hours at two intervals. The second data set is from the Deep Space Climate Observatory (DSCOVR; Burt and Smith 2012) Mission launched in early 2015. This spacecraft has been collecting high-spatial resolution images of Earth from Lagrange point 1 for approximately a decade. The Earth Polychromatic Imaging Camera (EPIC) on board DSCOVR covers the UV-NIR in 10 narrow spectral bands (0.6-3 nm width) and is specifically designed to address critical climate aspects such as the abundance of O3, SO2, aerosols, vegetation and clouds (Marshak et al. 2018). The spectral responses of the different wavelength bands have been calibrated, which enables direct monitoring of the Earth's reflectivity, be it at the narrow intervals of the instrument (Herman et al. 2018, Marshak et al. 2018, Li et al. 2022). In this section, the observations from the DSCOVR spacecraft are selected as the reference data that the other sources are compared to. In the following analysis, we compare results from several other studies to this full-disk, space based data to establish the most reliable reference spectrum for Earth.

### 5.2 Earth Data

**Earth Composite Spectrum:** Synthetic reflectance spectra are considered in this section from various radiative transfer simulations and validated with observations from space-based instruments. The photometric data from DSCOVR serve as the key reference spectrum for this section, and we scaled the other sources relative to it in Figure 4. Since we do not presently have reliable spectra recorded with space-based instruments over the entirety of the range of interest, we use simulated observations from the Planetary Spectrum Generator (PSG). The Kofman et al. 2024 simulated data are used as the output (composite) spectrum due to its ability to accurately reproduce spectral intensities from the DSCOVR satellite and cover the entirety of the desired spectral range. The resultant spectrum for Earth includes geometric albedo from 0.1 to 2.5 μm with a resolving power of 750, slightly higher than what was explored in Kofman et al., 2024.

The DSCOVR observations, as well as the Kofman et al., 2024 simulations, are the average over 24 hours on June 21st, 2022.

**L2011 Observations:** Observations in Livengood et al. 2011 were collected using the Deep Impact flyby spacecraft (Blume 2005), following the completion of a mission to Comet Tempel 1 in 2005. The Extrasolar Planetary Observation and Characterization (EPOCh) and Deep Impact eXtended Investigation (DIXI) scientific proposals for the extended mission of the Deep Impact spacecraft were combined to form EPOXI (Klaasen et al. 2013), that was initiated in June 2007. Data were collected using the High-Resolution Instrument (HRIVIS) with seven filters, centered at wavelengths between 372 nm and 948 nm, each with an approximate width of 100 nm. Additionally, infrared spectra were recorded with the HRIIR ranging from 1100–4540 nm. The HRIIR has a spectral resolving power ranging from 215-730 between 1.1-2.5 μm (reported resolving power of 730 near 1.1 μm and 215 near 2.5 μm). Data from EPOXI were collected over three 24-hour periods beginning on March 18, May 28, and June 4 2008. In this work the data are included from their first Earth observation, with a solar phase angle of 57.7° and a 77% illuminated disk of Earth. Data are scaled to phase zero in this work by multiplying the continuum by the Lambert phase coefficient of 1.582.

**R2011 Model:** Synthetic data from Robinson et al. 2011 are validated relative to the EPOXI Earth Data sets recorded with the High-Resolution Instrument on the Deep Impact flyby spacecraft. Data in this source are published as disk-integrated Earth reflectivity values. The model incorporates spatially resolved atmospheric composition and cloud and surface variability features across the globe. The three-dimensional model of Earth reproduces the temporal variability and absolute brightness in the visible, near and mid infrared (Robinson et al. 2011). Reflectance values are described in their work as π times the disk-integrated radiance ($W/m^2/\mu m/sr$) divided by the solar flux at 1 au (in $W/m^2/\mu m$) at phase zero. The EPOXI observations are taken at angles of 57.7° and 76.6° and the data are validated by running the model to match the observed phase angles from EPOXI. Since the data are initially recorded at a phase angle of zero, there are no geometric scale factors applied. The model provides Earth reflectivity data from 0.51 to 1.68 μm, and it is plotted in Figure 4.

**R2017 Model:** Roberge et al. 2017 provides geometric albedo spectra for each of the solar system planets (excluding Mercury). The spectrum is calculated from 0.3-2.5 μm using the Spectral Mapping Atmospheric Radiative Transfer (SMART) model (Meadows & Crisp 1996; Crisp 1997) and reflectance models from the Virtual Planet Laboratory. Notably, the model used in Roberge et al. 2017 was also used in Robinson et al. 2011 to create their Earth reflectivity spectra. In this work the planet is simulated as a single pixel of light detected by the instrument, providing disk-averaged data with a resolving power of 300. The model provides a spectrum of modern Earth with an equatorial view averaged over the 24-hour rotation period. In their work the spectrum is validated with the same EPOXI Earth observations described above in Robinson et al. 2011.

**D2022 Observations (photometric reference):** Data are included from the DSCOVR observations of the Earth's reflectivity as measured on the 21st of June 2022. The EPIC instrument onboard DSCOVR recorded Earth's reflected light intensity as CCD counts over 10 narrow spectral channels spanning 0.317-0.779 μm. The central wavelength and full width at half maximum (FWHM) values for each bandpass are defined in Marshak et al. (2018). Observations are recorded at 9 different times of day and the average geometric albedo values are included in this work for each spectral channel. Inclusion of observations across various times of day allows a more comprehensive reference for the reflected light spectrum of Earth, as it combines data across the heterogeneous surface features of the planet. The measurements are calibrated by correcting CCD counts to correspond with reflectance values of the nadir instruments MODIS onboard Aqua and Terra satellites in the near-visible and near-infrared, between 0.443 and 0.780 μm (Marshak et al. 2018) and to the Suomi National Polar-Orbiting Partnership Ozone Mapping and Profiler Suite (Suomi NPP–OMPS) in the UV, between 0.318 and 0.388 μm (Li et al. 2017; Herman et al. 2018). The process to convert data collected with EPIC in counts per second to reflectance values is discussed in detail in Marshak et al. 2018. The calibrated disk integrated data from DSCOVR provides the best absolute reference for validating modeled data of Earth and is used as the photometric reference spectrum in this section.

**K2024 Model:** These modeled Earth reflectance spectra were computed using the Planetary Spectrum Generator (PSG, https://psg.gsfc.nasa.gov, Villanueva et al. 2018, 2022), and are based on full-disk spectra, as reported in Kofman et al. (2024), validated with narrow band photometric observations from DSCOVR. The simulations leveraged the Modern-Era Retrospective analysis for Research and Applications (MERRA-2) 3D climate database, which reports a large range of atmospheric variables at high temporal and spatial resolutions. By ingesting the 3D distribution of temperature, pressure, clouds, and several spectroscopically relevant molecules into the Global Emission Spectrum (GlobES) module of PSG, the atmospheric profile was reproduced in Kofman et al. 2024. In addition to the atmospheric parameters, surface-cover types were adopted from the Moderate Resolution Imaging Spectroradiometer (MODIS), providing a limited set of 5 types of surfaces reflectance that were based on reflectance values from the United States Geological Survey (USGS). The reference spectrum was created by taking the average of 9 different reflectance spectra, corresponding to interval time steps over a 24-hour period of the Earth. The resulting disk averaged spectrum is shown in Figure 4.

# 6 Mars

## 6.1 Mars Introduction

Mariner IV was the first spacecraft to observe Mars in close proximity during its flyby, reaching its closest approach with the planet on July 15th, 1965. Images from this mission were historic and marked the first time that another planet was imaged from deep space. A key finding of this mission was that Mars topography was heavily cratered, resembling the moon more closely than the Earth (Abelson 1965). Initial findings of a weak magnetic field, lack of a radiation belt, and a tenuous atmosphere led to the important realization that Mars and Earth host very different planetary environments and likely had different histories. Mars is now the most visited planet in our solar system and has been the target of many different flybys, orbiters, landers and rovers. It currently hosts 2 active rovers on the surface and 7 orbiting satellites. In the 60 years since Mars was first visited our understanding of this nearby planet has grown immensely. In this work we consider reflectance spectra of Mars from a combination of ground and space-based instruments. It is important to note that the surface of Mars varies greatly in composition, in brightness and in spectral morphology, and therefore the here defined “photometric” reference should be understood as a representative spectrum of Mars, but the spectrum would vary depending on the sampled regions on the planet. In the following sub-sections, we describe each of the sources used to create the final plots presented in Figure 5 and the corresponding scale factors are recorded in Table 4.

## 6.2 Mars Data

**Mars Composite Spectrum:** True and synthetic reflectance spectra are considered in this section from various sources. The Mallama et al. 2017 photometric data serve as the reference spectrum for Mars, and the observations from the other sources are scaled relative to it in Figure 5C. As with Earth, we do not presently have reliable spectra recorded with space-based instruments over the entirety of the range of interest, so we used simulated observations from the Planetary Spectrum Generator (PSG). The PSG simulated data are used for the output (composite) spectrum since they could accurately reproduce spectral intensities from the sources described below. The resultant spectrum for Mars includes geometric albedo data from 0.30 to 2.45 μm with a resolving power of 500.

**MW1971 Observations:** Reflectance spectra from McCord and Westphal 1971 were collected with the 60-inch telescope and double beam photometer at the Cerro Tololo Inter-American Observatory. Data are included from their observations on May 25 and 27, 1969. The phase angle at the time of observations was only 5°, and multiple regions on the visible disk were selected to approximate the geometric albedo in their work. Localized albedo values were published for seven different regions on Mars, each with a diameter of 200 km. In this work we included albedo spectra for the darkest (Syrtis Major) and brightest (Arabia) regions on the surface. The Syrtis Major data are referred to as MW1971_SM and the Arabia data are called MW1971_A and displayed in Figure 5. In Figure 5C, MW1971_SM is multiplied by an empirical scale factor of 1.7 and MW1971_A is multiplied by 0.7 to best align each spectrum to the reference data from Mallama et al. 2017 (see below M2017).

**W1972 Observations:** Data in Wallace et al. 1972 were collected with the Wisconsin Experiment Package onboard the second Orbiting Astronomical Observatory (OAO-2) (Code et al. 1970), following its launch in December 1968. Observations were recorded on April 23, 1969, and albedo values were presented spanning 0.204 to 0.365 μm. Mars was observed at a phase angle of 26.8° and is adjusted to a phase angle of zero using the Lambert phase integral equation and applying a geometric scaling factor of 1.107 in Figure 5B.

**M2017 Observations (photometric reference):** Geometric albedo values are determined for Mars within the Johnson Cousins photometric system from observations in Mallama et al. 2017 and the references therein. The photometric data span a range of 0.36 to 0.90 μm containing 7 different bandpasses adopted from the earlier Mallama 2007 publication. The data are displayed in Figure 5 with horizontal bars representing the filter width for each of the values, centered at the effective wavelengths for each Johnson Cousins band.

**R2017 Model:** The model spectrum is generated from 0.3-2.5 μm using the Spectral Mapping Atmospheric Radiative Transfer (SMART) model (Meadows & Crisp 1996; Crisp 1997) and considering reflectance models from the Virtual Planet Laboratory (VPL). The planet is simulated as a single pixel of light detected by the instrument, providing disk-averaged data with a resolving power of 300.

**PSG Model:** The model Mars spectrum was generated with PSG using current standard Mars atmospheric conditions, including molecular absorptions and Rayleigh scattering for CO2, H2O and O3 and dust and water-ice clouds scattering. The surface was modelled assuming a Lambertian surface considering reflectance spectra from the Compact Reconnaissance Imaging Spectrometer for Mars (CRISM) on Mars Reconnaissance Orbiter (MRO) (Murchie et al. 2007, Seelos et al. 2024). The simulation was computed with a resolving power of 500 employing PSG's multiple scattering solver (PSGDORT) considering 2 pairs of scattering streams. Data span 0.200 to 2.499 μm and are used as the output (composite) spectrum for Mars.

# 7 Jupiter

## 7.1 Jupiter Introduction

Jupiter is the most massive planet in our solar system, with a mass approximately 318 times that of Earth. It is dominated by a thick atmosphere and has a relatively low density of 1.3 g/cm$^3$, lower than that of each of the terrestrial planets. Its atmosphere is composed primarily of hydrogen and helium, with traces of heavier elements. While its harsh environment is unlikely to support life, studying Jupiter provides valuable insights into the origin of our solar system. The abundance of light elements suggests that Jupiter likely formed early in the solar system's history, when the protoplanetary disk was still rich in gas (Lunine et al. 2004). Jupiter has been observed extensively through ground-based telescopes and has been the target of multiple space missions. Pioneer 10 was the first spacecraft to journey beyond Mars, pass through the asteroid belt, and study Jupiter up close. It reached the planet and began taking the first observations while in orbit on December 4th, 1973. Among the key findings were direct measurements of the magnetic field and the composition of the upper atmosphere. The planet was then studied in detail by Cassini (Jaffe et al. 1996, Spehalski et al. 1996, Matson et al. 2002) during a flyby en route to Saturn. Extensive data were collected between October 1, 2000 and March 22, 2001. The Cassini mission to Jupiter focused on studying the 3-D atmospheric structure, global energy balance, and magnetosphere, among various other scientific objectives (Matson et al. 2002). Cassini's study of Jupiter was impactful to the understanding of clouds, hazes, aerosols, and the evolution of cloud structures and winds that describe the atmospheric dynamics (Porco et al. 2003). Reflectance spectra from Cassini instruments published by Coulter et al. (2022) and Li et al. (2018) are incorporated in this work. NASA's Juno spacecraft was launched on August 5th, 2011 and entered orbit around Jupiter on July 4th, 2016. The science goals for this mission were broadly focused on understanding the origin and formation of the planet and did so by specifically targeting the thermal evolution and interior structure, the internal magnetic field, and dynamical processes in the atmosphere and magnetosphere (Bolton et al. 2017). Observations from the Juno mission published by Giles et al. (2022) are included in this work, and beyond these, we also consider reflectance spectra of Jupiter from a wide range of ground-based observations.

### 7.2 Jupiter Data

**Jupiter Composite Spectrum:** The ground and spaced based observations in this section are combined to produce a composite spectrum for Jupiter, spanning 0.123 to 2.491 μm. The composite spectrum stitches reflectance spectra from the International Ultraviolet Explorer published in Clarke et al. 1982 [0.123- 0.180 μm], Cassini observations published Li et al. 2018 [0.180-0.300 μm], European Southern Observatory observations published Karkoschka 1998 [0.300-1.037 μm], and Cassini observations from Coulter et al. 2022 [1.041-2.491 μm]. In our study, the results from Karkoschka 1998 serve as the key reference spectrum, and we scaled the other sources relative to it. The composite spectrum contains scaled versions of the original data, using the factors described below and in Table 5.

**C1979 Observations:** Albedo spectra from Clark and McCord 1979 were collected using the 2.24 m telescope at the Mauna Kea Observatory. Apertures were focused at both the center of the disk and the east limb of the planet, and data were separately collected for each location. The final spectra published in their work are the average of 8 observing runs for the center and 6 runs for the limb position, all recorded on November 4, 1976. Data are included from 0.661-2.504 μm for the center aperture and from 0.661-2.501 μm for the limb. In this section the center data are labeled as C1979_c, while the limb data are referred to as C1979_l. In Figure 5B the data recorded at the center is multiplied by a correction factor of 0.66 to account for Lambertian limb darkening of the visible disk of the planet.

**C1982 Observations:** Geometric albedo data were sourced from Clarke et al. 1982, with observations made using the International Ultraviolet Explorer (IUE; Boggess et al. 1978). From December 1978 to June 1979 reflected light was observed from 0.119-0.175 μm. A composite spectrum was created for this range by summing eleven 15-minute exposures. Additional IUE observations were conducted from May to July 1980, spanning 0.170 to 0.195 μm, and a second spectrum was produced by summing three 5-minute exposures. The results from each of the spectral regions were combined to produce the final composite spectrum in their publication and included in this work. No scale factors were applied to this dataset, which includes data from 0.123-0.194 μm. These data are included in the composite spectrum for Jupiter.

**K1998 Observations (photometric reference):** Jupiter was observed using the European Southern Observatory's 1.52 m telescope, and results were published in Karkoschka 1998. A full-disk albedo spectrum, ranging from 0.300 to 1.050 μm, was produced using data collected between July 6-10, 1995. The data were recorded with a spectral resolution of 0.4 nm between 0.300-0.995 μm and 1 nm between 0.995-1.050 μm. Their publication improves upon previous work in Karkoschka 1994 by observing Jupiter at low air mass values of 1.1 or lower. The publication presents the full-disk albedo as I/F data averaged over the planet's disk, taken at a specific phase angle, while geometric albedo refers to the same value at phase zero. The planet was observed at a phase angle of 6.8° on July 7, 1995, and it is adjusted to a phase angle of zero using the Lambert phase coefficient of 1.007 in Figure 6B. This source is used as the reference spectrum for scaling and comparing the other sources to in the final plot displayed in Figure 6C. This spectrum was selected as reference due to the successful validation of the data across a broad range of instruments, precise calibration, and the high spectral resolution data that was produced. These data are included in the composite spectrum for Jupiter.

**B2003 Observations:** Reflectance data from Betremieux et al. 2003 were collected with the Faint Object Spectrograph (FOS; Ford 1990) onboard the Hubble Space Telescope. A 780 second exposure captured the Jovian equatorial region with FOS on June 5, 1993, and presented as an I/F continuum. The spectra span from 0.176 to 0.231 μm with a high resolution of 0.3 nanometers. The data reduction is described in detail in Betremieux and Yelle (1999).

**R2017 Observations:** Geometric albedo observations are included from Roberge et al. 2017. The geometric albedo continuum of Jupiter is created using data from Karkoschka 1998 and the SpeX instrument on the NASA Infrared Telescope Facility (Rayner et al. 2003). Here we include the portion of their data set that corresponds to the observations made with the SpeX instrument, spanning 1.002 to 2.497 μm. In their work, the flux densities were divided by the solar spectrum to create reflectance spectra and scaled to align with the continuum from Karkoschka 1998. It is also mentioned that there is a gap of 0.06 μm at approximately 1.85 μm in the spectrum that is patched using model giant planet spectra from Burrows et al. 2004. Since the region containing synthetic data is not clearly specified, the spectrum between 1.79 and 1.91 μm is displayed with a dashed line.

**L2018 Observations:** Observations were made with the Cassini spacecraft during the Jupiter flyby en route to Saturn, and results published in Li et al. 2018. The reflected solar radiance is measured with the Cassini Visible and Infrared Mapping Spectrometer (VIMS; Brown et al. 2004) and Imaging Science Subsystem (ISS; Kahn and King 1996) from October 2000 to March 2001. The first data set taken from this source contains monochromatic geometric albedo data from 0.015 to 3.990 μm. A scale factor of 0.87 was applied in Figure 6C to best align the continuum with the reference continuum from Karkoschka 1998. The monochromatic geometric albedo spectrum is displayed in dark orange in Figure 6. This source also compares their observations from Cassini ISS to the full-disk albedo spectrum in Karkoschka 1998. The Cassini full-disk albedo spectrum in Li et al. 2018 is measured at a phase angle of 10.3˚ and is included in this work from 0.359 to 1.018 μm. In Figure 6B, the dataset is adjusted to a phase angle of zero by multiplying the phase coefficient of 1.016 for a Lambertian sphere. The Cassini ISS data is included in the composite spectrum for Jupiter.

**G2021 Observations:** Reflectance data were collected with the Ultraviolet Spectrograph (UVS; Gladstone et al. 2014) on NASA's Juno Mission (Bolton et al. 2017), and the results published in Giles et al. 2021 are considered in this work. In their study, the reflectance is defined as the observed spectrum divided by the incident solar spectrum, corrected for the solar zenith angle (Giles et al. 2021). Dayside observations from August 2016 to November 2020 are focused on the equatorial region, specifically avoiding the auroras in the polar regions of the planet. Combined data produce the reflectance spectrum used in this work, from 0.14 to 0.20 μm. The spectrum already aligns well with the reference continuum from Karkoschka 1998, and no additional scale factors are applied.

**C2022 Observations:** Disk integrated I/F data are provided in Coulter et al. 2022. Observations were made using the Cassini (Jaffe et al. 1996, Spehalski et al. 1996, Matson et al. 2002) Visual and Infrared Mapping Spectrometer (VIMS) (Brown et al. 2004) instrument. The detector provides data from 0.5 to 5.2 μm, however only a subset of this data is included in this study. Data were limited to images where the entire disk of the planet was visible, none of the pixels were oversaturated, and no stray light sources were affecting the spectra. This reduced the image cubes available from approximately 14,000 to a few dozen, and finally 6 were chosen in their study. Images were taken during Cassini's Jupiter flyby between December 2000 and January 2001. The study reports phase angles from 1.66 – 133.52°, but in this work only the data recorded at 1.66° are included. Typically, a scale factor would be applied, however, since the data are so close to phase zero initially that the scale factor (1.0004) is negligible. In Figure 6C, the data are scaled by 1.186 to align with the visual geometric albedo at 0.55 μm, presented in Mallama et al. 2017. These data are included in the composite spectrum for this section.

## 8 Saturn

### 8.1 Saturn Introduction

Saturn is roughly one third of Jupiter's mass, however it is still the second most massive planet in our solar system, it is also similar to Jupiter in structure and dominated by its thick atmosphere. This planet is the least dense in the solar system, as it is composed nearly entirely of hydrogen and helium. With an average density of approximately 0.7 g/cm$^3$, it is also the only planet in our solar system that is less dense than water. Saturn is uniquely defined by its famous rings and nearly 150 moons, the most of any planet in the solar system. The rings of Saturn are composed predominantly of water ice, however the slightly reddened appearance is due to trace elements acquired during meteoroid bombardment (Cuzzi and Estrada 1998). Pioneer 11 was launched on April 6, 1973 and became the first spacecraft to closely approach the Saturn system, with its closest approach on September 1, 1979. Radio occultation measurements of the ionosphere and upper neutral atmosphere were made in the planet's southern hemisphere during the flyby, refining temperatures in the exosphere, revealing a main ionospheric peak at ~1,800 km and mapping ionosphere densities, defining the extended ionization to ~30,000 km, and estimating temperature and pressure limits of the neutral atmosphere with a principal minimum of 88 ± 4 K at 74 mbar (Kliore et al. 1980). Additionally, it provided the first direct detection of the presence of Saturn's magnetic field (Acuña and Ness 1980, Acuña et al. 1980, Smith et al. 1980). By the time Pioneer 11 reached the planet, the next mission was already launched and in progress. The Voyager 1 spacecraft made its closest approach on November 12, 1980 and confirmed and refined the findings from Pioneer 11's study of the magnetosphere and magnetic field (Ness et al. 1981). Voyager 1 returned photographs that revealed new details regarding the Saturn system, including Titan's haze, Saturn's unique cloud circulation patterns, subcomponents of Saturn's rings, and multiple new moons (Smith et al. 1981). The Cassini-Huygens Mission (Matson et al. 2002) is the most recent mission and the first to study Saturn in close proximity for a long duration. It launched on October 15, 1997 and entered the Saturn system in mid 2004 where it studied the planet through its changing seasons and monitored its composition for approximately 13 years. In this work we consider reflectance spectra of Saturn as collected from Cassini as well from a wide range of ground-based observations and space-based missions.

### 8.2 Saturn Data

**Saturn Composite Spectrum:** Using a variety of publications and data from both ground- and space-based instruments, we have constructed a composite spectrum of Saturn spanning 0.122 to 2.490 μm. In this section, the Karkoschka 1998 data serve as the main reference spectrum that each of the other sources were scaled to. In the composite spectrum, UV data from the International Ultraviolet Explorer, published in Clarke et al. 1982 is included from 0.122 to 0.190 mic μm rons. There is a small gap in the data set, then data recorded at the European Southern Observatory and published in Karkoschka 1998 are included from 0.300 to 1.050 μm.

Lastly, data from the Kitt Peak National Observatory, originally published in Fink and Larson 1979, is used for the continuum between 1.050 and 2.490 μm. The composite spectrum contains scaled versions of the original data, using the factors described below and in Table 6.

**IL1971 Observations:** Photometric observations of Saturn were published in Irvine and Lane 1971. The planet was observed at both Le Houga Observatory in France using a 12-inch Newtonian reflector and the Boyden Observatory using a 16-inch Cassegrain reflector in South Africa with identical auxiliary electrons and photometers. Data were recorded using 10 narrow band filters from 0.315 to 1.06 μm between 1963 and 1965. This study expands on a series of earlier publications that provide the brightness of the disk of the planet and the rings (Young and Irvine 1967; Irvine et al. 1968a, Irvine et al. 1968b) and it attempts to isolate the disk brightness. All of the data in their work has been adjusted to a phase angle of zero and the rings were corrected to a zero-inclination angle to provide the geometric albedo values. There are no geometric scaling factors applied; however the continuum is lower in intensity than the Karkoschka 1998 (see below K1998) reference spectrum, and it is therefore multiplied by an empirical scaling factor of 1.15 in Figure 7C.

**C1979 Observations:** Albedo spectra from Clark and McCord 1979 were collected using the 2.24 m telescope at the Mauna Kea Observatory. Apertures were focused at both the center of the disk and the west limb of the planet, and data were separately collected for each location. The final spectra published in their work are the average of four observing runs for each position, recorded on February 2, 1977. Data are included from 0.664-2.503 μm for the center aperture and from 0.673-2.504 for the limb. In this section the center data are labeled as C1979_c, while the limb data are referred to as C1979_l. In Figure 7B the data recorded at the center has a correction factor of 0.66 applied to account for Lambertian limb darkening of the visible disk of the planet. An additional scale factor of 1.05 was applied in Figure 7C to align the continuum with the reference full-disk albedo data in Karkoschka 1998.

**FL1979 Observations:** Spectral observations from Fink and Larson 1979 were collected using the Kitt Peak National Observatory's 4 m telescope and rapid scanning Michelson interferometer. Data were collected over three observing nights from February 28 to March 2, 1979. These observations were reduced to a resolution of 46 $cm^{-1}$ in their work and cover a range of 0.96 to 2.49 μm. Since the values are already converted to geometric albedo values in the original work, there are no geometric scaled factors applied to this data set. In Figure 7C the data are multiplied by a scale factor of 1.2 to align it with the continuum from Karkoschka 1998 (see below K1998). Data from this source are included in the composite spectrum for Saturn.

**C1982 Observations:** Geometric albedo measurements are included from Clarke et al. 1982, recorded using the International Ultraviolet Explorer (Boggess et al. 1978). Observations of Saturn were conducted between March and July 1980, covering wavelengths from 0.122 to 0.190 μm. The final disk-averaged continuum combines the highest quality spectral data from multiple observing runs across 3 wavelength bands (see source for details). Measurements were taken at the center of Saturn's disk, and the data are multiplied by a scale factor of two thirds to account for Lambertian limb darkening. Due to the light subtraction methods in their analysis, data between 0.175 and 0.190 μm are uncertain in magnitude and should be considered upper limits for the geometric albedo. Data from this source are included in the composite spectrum for Saturn.

**K1998 Observations (photometric reference):** Saturn was observed using the European Southern Observatory's 1.52 m telescope, and results were published in Karkoschka 1998. This source presents updated results from the original Karkoschka 1994 publication, and much of the analysis is conducted with the same methods as the earlier publication. The updated work includes observations of Saturn in a rare edge-on geometry, causing the rings to appear around 1000 times darker than usual and allowing for a more accurate determination of the planet's albedo value. The full-disk albedo presented in their work refers to I/F data averaged over the planet's disk, taken at a specific phase angle. The planet was observed at a phase angle of 5.7° on July 7, 1995, and a correction factor is applied to the continuum to adjust the data to a phase angle of zero degrees, providing the true geometric albedo. This spectrum spans 0.300 to 1.050 μm, with a spectral resolution of 0.4 nm between 0.300 – 0.995 μm and 1 nm between 0.995 – 1.050 μm. The continuum presented here has been successfully validated across a wide range of instruments and publications and is considered to be highly accurate due to the precise calibration and high spectral resolution data produced. This source is used as the reference spectrum that the other sources in this section are scaled to and is included in the composite spectrum for Saturn.

**R2017 Observations:** Like Jupiter, the spectrum for Saturn from Roberge et al. (2017) was created using data from Karkoschka 1998, the SpeX instrument on the NASA Infrared Telescope Facility (Rayner et al. 2003), and model data (Burrows et al. 2004). Here we include the portion of the R2017 data set that corresponds to the observations made with the SpeX instrument, spanning 1.002 to 2.497 μm scaled in their work to align with the continuum from Karkoschka 1998 (K1998).

**C2022 Observations:** Disk integrated I/F reflected light spectra is provided in Coulter et al. 2022. Observations were made using the Cassini (Jaffe et al. 1996, Spehalski et al. 1996, Matson et al. 2002) Visual and Infrared Mapping Spectrometer (VIMS) (Brown et al. 2004). The detector provides data from 0.35 to 5.13 μm, however only a subset is included in this study, ranging from 0.35 to 2.5 μm. Their analysis only includes images where the entire disk of the planet was visible, none of the pixels were oversaturated, and no stray light sources were affecting the spectra. This reduced the image cubes available from approximately 14,000 to a few dozen, and finally 6 were chosen in their study. The study reports phase angles from 39.6-110.2˚, but in this work only the observations recorded at 39.6˚ were included. The spetra included here were recorded with Cassini on August 2, 2007 and are multiplied by a scale factor of 1.24 in this work to adjust the data to a phase angle of zero in Figure 7B. In Figure 7C, the continuum is scaled again with a scale factor of 1.66 to best align with the reference continuum from K1998.

## 9 Titan

### 9.1 Titan Introduction

Similar in size to Mercury, Titan is Saturn's largest moon and the second largest in the solar system. It is the only known planetary satellite to host a substantial atmosphere, making it a unique scientific target. Titan's atmosphere is dense and complex, and even displays evidence of liquid surface erosion and high latitude lakes (Lunine et al. 2008). Of all the moons in the solar system, Titan contains the most Earth-like atmosphere, consisting of 95% nitrogen, 4% methane, and 1% trace species at the surface, and a pressure of only 1.5 bars (MacKenzie et al. 2021). Clouds on Titan have been studied extensively (see Rodriguez et al. 2009 and references therein) and form from the condensation of methane and ethane. With less radiant energy and a more massive atmosphere, the weather on Titan is relatively calm and "sluggish" compared to Earth. Titan receives about 100 times less power from the sun, resulting in immensely cold surface temperatures of approximately 94 K (Griffith et al. 2000). Methane on Titan plays an analogous role to water on Earth, existing in liquid form on the surface and forming precipitation cycles with clouds and rain (Atreya et al. 2006, Lunine and Atreya 2008). Titan's atmosphere also supports the formation of organic hazes in the presence of oxygen-bearing molecules (Horst 2017), presenting an intriguing opportunity to investigate prebiotic chemistry that may be linked to the origins of life. Titan has been visited by various space-based missions, beginning with the Pioneer 11 spacecraft. This spacecraft launched on April 5, 1973, and was the first mission to venture beyond the Kuiper belt and encounter the Saturn system in 1979. Directly following Pioneer 11, the Voyager space probes approached Titan as part of their exploration of the Saturn system. In 1980, Voyager 1 conducted a broad study of Saturn and its moons, including a very close flyby of Titan at approximately 4,000 miles. Radio signals from Voyager 1 provided new information on the atmosphere and radius of Titan (Lindal et al. 1983). Radio occultation measurements confirmed that molecular nitrogen is the dominant atmospheric constituent, advanced the understanding of potential cloud types, and suggested methane's important role in Titan's atmosphere after revealing its capability to exist in three phases (Tyler et al. 1981). In 1997, the NASA Cassini Mission (Matson et al. 2002) was launched and arrived at the planet in June 2004. It included ESA's Huygens Titan Probe that descended into Titan's upper atmosphere in January 2004 and is still the only probe to land on Titan. Cassini-Huygens revolutionized our understanding of Saturn's moon and transformed it from a hazy orange blur to a detailed world when it took the first images from the surface. Its precise in situ measurements also revealed new and more detailed information regarding the atmospheric composition and profile (Niemann et al. 2005).

Titan's atmosphere is particularly valuable to this study as it provides the reflectance signature of a planet still under investigation for potential habitability. It is indeed a strong candidate for astrobiological studies due to its similarities with Earth; the methane cycle is analogous to Earth's water cycle, and its active organic chemistry includes compounds relevant to prebiotic processes (see Raulin 2008 and references therein). Developing a strong reference for Titan's atmosphere and reflected light spectrum enhances our ability to better model exoplanets that may be similar to Titan in nature. In 2018, Riba et al. (2018) have announced the detection, via radial velocity measurements, of a large rocky and non-transiting exoplanet beyond the snowline called Barnard's star b. This planet was postulated to be the first super-Titan, a large rocky exoplanet with equilibrium temperature comparable to that of Titan. Owing to the very close proximity of the Barnard's star, the second closest system from Earth, Felton et al. 2020 examined how the LUVOIR coronagraph ECLIPS could be able to spatially resolve the planet from the star and to spectrally analyze its atmosphere (assumed to be Titan-like). This study concluded that potentially Titan-like exoplanets would be characterizable with a large space-based telescope. Unfortunately, the detection of Barnard's star b has then been refuted a few years later (Lubin et al. 2021). Nevertheless, that kind of study extrapolates what we know from Titan to predict observables on an exoplanet, having accurate spectral data for Titan along with realistic atmospheric models and data is paramount. This section incorporates data from the Cassini mission and include spectroscopic sources from several other ground-based and space observatories.

### 9.2 Titan Data

**Titan Composite Spectrum:** Using a variety of publications and data from both ground- and space-based instruments, we have constructed a composite geometric albedo spectrum of Titan spanning 0.180 to 2.495 μm. In this section, the Karkoschka 1998 data serve as the main reference spectrum that each of the other sources were scaled to. In the composite spectrum, data from the Hubble Space Telescope, published in McGrath et al. 1998 are included between 0.180 and 0.330 μm. Observations from the European Southern Observatory are included from Karkoschka 1998, between 0.330- 1.040 μm. Observations from the Canada France Hawaii Telescope are included from Negrao et al. 2006 [1.053-1.348 μm, 1.520-1.805 μm, 1.965-2.495 μm] and the gaps are filled in using synthetic data from Negrao et al. 2005 [1.350-1.515 μm, 1.810-1.965 μm μm]. The composite spectrum contains scaled versions of the original data, using the factors described below and in Table 7.

**Y1974 Observations:** Photometric geometric albedo data are considered from Younkin 1974, covering a range of 0.50 to 1.08 μm. Data were collected at the Mount Wilson Observatory on January 2 and 3, 1974 with a 3 nm bandpass using the 60-inch reflector (Ritchey 1909), and Fastie-Ebert spectrometer. An empirical scale factor of 0.8 is applied in Figure 8C to align the continuum with recorded values from Karkoschka 1998 (see below K1998). Data are displayed in Figure 8 with small round symbols and connected with a dotted line for visual continuity.

**N1984 Observations:** Titan was observed with the McDonald Observatory's 2.1-m telescope and the ES-2 spectrograph on May 17-18, 1981, and results are included from Neff et al. 1984. The planet was observed at a phase angle of approximately 4.9˚ on each night. Monochromatic flux ratios were determined in Neff et al. 1984 and used to compute the monochromatic geometric albedo (Neff et al. 1984, Equation 1). Data are published in units of geometric albedo in this source, however, they were not presented at a phase angle of zero, so this source presents a close approximation of the geometric albedo. The values were published in their work from 0.350-1.050 μm at a spectral resolution of approximately 0.7 nm. The authors mentioned that the phase angles were not adjusted to a phase angle of zero because the phase dependency to both angle and wavelength were not well determined at the time. In this work, we calculate a Lambert phase correction of 1.004, using Equation 1, and multiple values in the continuum by the correction factor in Figure 8B. Neff et al. 1984 expands on measurements obtained by Younkin 1974, and provides the higher resolution and signal to noise ratio data required for a detailed quantitative analysis of the reflected light spectrum of Titan. The geometric albedo data agrees well with the Karkoschka 1998 spectrum, and no scale factors are applied.

**K1998 Observations (photometric reference):** Titan was observed using the European Southern Observatory's 1.52 m telescope, and results were published in Karkoschka 1998. A full-disk albedo spectrum, ranging from 0.300 to 1.050 μm, was produced using data collected between July 6-10, 1995. The data were recorded with a spectral resolution of 0.4 nm between 0.300-0.995 μm and 1 nm between 0.995-1.050 μm. Titan was observed at a phase angle of 5.7° on July 7, 1995, and the data were scaled to a phase angle of zero in this work using the corresponding Lambert phase integral equation, of approximately 1.005. The publication presents the full-disk albedo as I/F data averaged over the planet's disk, taken at a specific phase angle, while geometric albedo refers to the same value at phase zero. For this study, reflected light spectra from 0.305-1.050 μm are included to avoid an anomalous spike in the data near the edge of the spectrum. This source is used as the reference spectrum for scaling and comparing the other sources because of the successful validation across a broad range of instruments, precise calibration, and high spectral resolution data. Data from this source are included in the composite spectrum for Titan.

**M1998 Observations:** Ultraviolet albedo measurements were made using the Faint Object Spectrograph (FOS) (Keyes et al. 1995) on the Hubble Space Telescope (Chaisson and Villard 1990, Endelman 1991) on October 9, 1991 and August 25, 1992 and presented in McGrath et al.

1998. Data in October and August were originally observed with phase angles of 5.5° and 1.2° respectively and then corrected to a phase angle of zero using the equation in Lockwood et al. 1986. The adjusted disk-averaged geometric albedo data range from 0.18-0.33 μm. In their study, data are compared to observations from Karkoschka 1994, an earlier study collecting full-disk albedo spectra for the Jovian planets and Titan, and a strong agreement is observed between the two datasets. The continuum is scaled by an empirical factor of 1.2 in Figure 8C for consistency with the Karkoschka 1998 (K1998) full-disk albedo. Data from this source are included in the composite spectrum for Titan.

**N2006 Observations and Model:** Geometric albedo data are included from Negrao et al. 2006, recorded with the Fourier Transform Spectrometer (FTS) (Maillard and Michel 1982) on the Canada France Hawaii Telescope (CFHT). Multiple observations were reported in this reference and data was included in this work from 1994 and 1995. Data were recorded on September 23, 1994, and included from 0.872 to l.346 μm, 1.434 to 1.785 μm, and 1.963 to 2.499 μm. Additionally, data recorded on August 17, 1995, are included from 1.040 to 1.348 μm, 1.520 to 1.805 μm, 1.963 to 2.495 μm, with discontinuities between each of the described spectral regions. The 1995 data was recorded within three spectral bands between 1.040 and 2.495 μm. Data included from the J band span 1.040 to 1.348 μm, the H band from 1.520 to 1.805 μm, and the K band from 1.963-2.495 μm. The spectral resolutions for the J, H, and K bands are 0.52 nm at 1 μm, 0.66 nm at 1 μm, and 0.87 nm at 1 μm respectively, as reported in the original source. The average resolving powers were calculated here for the J, H, and K bands to be 2298, 2492, and 2598, respectively. Discontinuities exist between each of the described spectral bands and synthetic data from this source is included to provide a complete continuum. This source incorporates a radiative transfer code from Rannou et al. 2003 combined with methane absorption coefficients from Irwin et al. 2006 to provide the simulated geometric albedo spectrum of Titan between 1.050 and 2.490 μm. The synthetic data are compared and adjusted to the CFHT data presented in the same source to best match the model to recorded data for Titan. The synthetic data are plotted with a light blue dashed line filling in the gaps in data in Figure 8 and labeled as N2006_m (model). In the composite spectrum, only the 1995 data and model are included.

**L2008 Model:** Synthetic geometric albedo data are presented in Lavvas et al. 2008. Data covering a range from 0.153 to 5.930 μm are well validated against various other observations (Lavvas et al. 2008). The authors concluded that the model matches many of the features detected in the Cassini/Huygens observations and reproduces well the vertical profiles for the observed species. Their 1D model incorporates a variety of complex atmospheric processes, considering radiative/convective, photochemical, and microphysical interactions, to study Titan's haze and chemical species formed in the atmosphere. To best match the simulated data with the reference continuum, from Karkoschka 1998 (K1998), the data are multiplied by a scale factor of 1.22 in Figure 8C.

## 10 Uranus

### 10.1 Uranus Introduction

Uranus' semi-major axis is approximately 19.2 astronomical units from the sun, and along with Neptune, is classified as an "ice giant". It is the least massive of the four giant planets with a mass of approximately 14.5 times that of Earth, and it is the second least dense planet overall with an average density of 1.27 g/cm$^3$. Uranus orbits at the strongest tilt of all of the planets, with an axial tilt of 98° it appears to rotate on its side. Over the course of a season on Uranus one of the poles will experience 21 years of direct solar illumination while the other experiences complete darkness. The atmosphere is unique in our solar system due to the low flux and seasonal extremes due to the high obliquity of the planet. Giant planets are believed to form with nearly zero obliquity and the planets current tilt has been under investigation for various possible scenarios such as the tilt developing during planetary migration with the presence of an additional satellite (Boué and Laskar 2010) or a giant impact scenario earlier in its formation (Esteves et al. 2025 and references therein) and the possibility of many smaller collisions resulting in the high obliquity (Rogoszinski and Hamilton 2021). However, many questions regarding the characteristics of Uranus remain unanswered due to its low brightness and the challenges associated with reaching one of our most distant satellites. Voyager 2 is the only spacecraft to observe the planet in close proximity. The spacecraft was occulted by Uranus on January 24th, 1986 and the radio link between the planet and Earth was analyzed after passing through the atmosphere of Uranus at latitudes 2° and 7° S. The profiles in height of the ionosphere and gas refractivity, number density, pressure, temperature, and methane abundance in the troposphere and stratosphere were determined with results from Voyager 2 (Lindal et al. 1987). Voyager 2 also detected an intrinsic magnetic field and magnetosphere at Uranus (Ness et al. 1986), and nearly 7000 images of the planet's southern hemisphere, rings, and its satellites were collected, revealing cloud patterns from methane condensation, information about the zonal winds, detailed features in the rings, new satellites, and variable crater populations amongst the planet's satellites (Smith et al. 1986). Since this mission, the planet has only been observed from afar, however the recent decadal survey (National Academies of Sciences, Engineering, and Medicine 2021) recommended that NASA pursues a major new mission to Uranus this decade. An important motivation for the increased interest is that approximately 30 percent of all known exoplanets are ice giants, further incentivizing a deeper investigation of one of our least known planetary systems. The ice giants provide important insight to understand planets with hydrogen atmospheres. Detailed exploration of Uranus in upcoming missions can provide insight into transport processes in hydrogen atmospheres, constraints on the structure, interior composition, dynamo of Uranus, and insight that provides keys to understanding the origin of the Solar System (Guillot 2022). In this work we consider reflectance spectra of Uranus from a combination of ground and space-based instruments.

### 10.2 Uranus Data

**Uranus Composite Spectrum:** Using a variety of publications and data from both ground- and space-based instruments, we have constructed a composite geometric albedo spectrum of Uranus spanning 0.222 to 2.473 μm. In this section, the Karkoschka 1998 data serve as the main reference spectrum that each of the other sources were scaled to. In the composite spectrum, data from the Hubble Space Telescope, published in Courtin 1999 are included between 0.222-0.320 μm. Observations from the European Southern Observatory are included from Karkoschka 1998, between 0.320- 1.000 μm. Additionally, observations from the Infrared Telescope Facility are included, published in Irwin et al. 2022 between 1.003 and 2.473 μm. The composite spectrum contains scaled versions of the original data, using the factors described below and in Table 8.

**L1983 Observations:** Observations from Lockwood et al. 1983 are collected using the 1.8 m Perkins telescope at the Lowell Observatory. Reflected light flux values were recorded on April 24, 1981, between 0.522 and 0.763 μm at a spectral resolution of 0.8 nm. Additional observations were collected between 0.332 and 0.547 μm at a spectral resolution of 0.4 nm on June 17, 1981. In their work, geometric albedo values were determined across each of the wavelength regions. The data are scaled to align with the reference continuum from Karkoschka 1998 (K1998), and multiplied by an empirical scale factor of 1.05 in Figure 9B.

**N1984 Observations:** Uranus was observed with the McDonald Observatory's 2.1-m telescope and the ES-2 spectrograph on May 17-18, 1981, and results are included from Neff et al. 1984. The planet was observed at a phase angle of approximately 0˚ on each night. Monochromatic flux ratios were determined in Neff et al. 1984 and used to compute the monochromatic geometric albedo (Neff et al. 1984, Equation 1). The values were published in their work from 0.350-1.051 μm at a spectral resolution of approximately 0.7 nm. The full-disk albedo data is multiplied by an empirical scale factor of 1.08 in Figure 9B to align with the reference continuum Karkoschka 1998 (K1998).

**W1986 Observations:** Observations in Wagener et al. 1986 were collected using the large aperture of the Long Wavelength Prime (LWP) camera on the International Ultraviolet Explorer (IUE) satellite. Uranus was observed several times on October 2, 1985, and data were calibrated and combined to produce geometric albedo values. Three different geometric albedo curves were produced in their work by dividing the planetary fluxes by the scaled and extinction corrected stellar fluxes of two solar analogs (16 Cyg A and B) and the 1980 solar spectrum from Mount & Rottman (1981, 1983). The results of the three curves were averaged to produce the final geometric albedo curve included in this work, spanning from 0.205 to 0.334 μm. The standard deviation of the combined data set is described in their work.

**K1998 Observations (photometric reference):** Uranus was observed using the European Southern Observatory's 1.52 m telescope, and results were published in Karkoschka 1998. A full-disk albedo spectrum, ranging from 0.300 to 1.050 μm, was produced using data collected between July 6-10, 1995. The data were recorded with a spectral resolution of 0.4 nm between 0.300-0.995 and 1 nm between 0.995-1.050. Titan was observed at a phase angle of 5.7° on July 7, 1995, and the data were scaled to a phase angle of zero in this work using the corresponding Lambert phase coefficient (Equation 1), of approximately 1.005. The publication presents the full-disk albedo as I/F data averaged over the planet's disk, taken at a specific phase angle, while geometric albedo refers to the same value at phase zero. This source is used as the reference spectrum for scaling and comparing the other sources to in the final plot displayed in Figure 9B an used in the composite spectrum for Uranus.

**C1999 Observations:** Reflected light spectra were recorded with the Hubble Space Telescope's Faint Object Spectrograph (FOS) in Courtin 1999 on June 22, 1992, as part of the GTO Program 1290-GTO/ OS-86G. The geometric albedo spectra were calibrated using flux spectra from a common solar spectrum measured by the Solar-Stellar Irradiance Comparison Experiment (SOLSTICE) (Rottman et al. 1993) on the Upper Atmospheric Research Satellite. The observations of the solar spectrum and the reflected light from the planet were taken approximately one year apart to include a higher resolution solar spectrum, most similar to the FOS instrument. This incurs additional calibration, and the data were compared with spectra from the International Ultraviolet Explorer and European Southern Observatory and corrected with the absolute scale in Karkoschka 1998 (K1998). Additional details on the calibration of the data are presented in C1999. Data from this source are included in the composite spectrum for Uranus.

**S2015 Observations:** Observations from Schmude et al. 2015 report on long observational campaigns of Uranus, spanning the course of 25 years. Furthermore, the work presents a detailed reanalysis of an observational series by Lockwood and Jerzykiewicz 2006, who also kept a long record of observations of the photometry of Uranus (1972-2016). Lockwood and co-workers report the brightness of the planet in the *b* and *y* bands (472 and 551 nm respectively) nearly continuously over 50 years; Schmude observed in the B, V, R and I bands (436 nm, 549 nm, 700 nm and 900 nm respectively). Although the compilation work by Schmude is not peer reviewed, it presents one of the most complete reviews of observational studies of Uranus. In both observational studies, the brightness of the planet is scaled to reference stars near the field of view, which is subsequently scaled to albedo by considering the distance and the size of the planet by the authors. Evaluated over the range of observed sub-latitudes, the short-wavelength (0.4 um) fluxes change by a modest 3% while the red (0.7 μm) fluxes vary by a more substantial 30%. These variations in the infrared primarily relate to changes in the sub-Earth/Sun latitudes, and the contributions to the poles, which are much brighter, to the hemispheric disk fluxes. We here present their average geometric albedo values, which are determined from the available observations within each of the available filter widths. This variability in the infrared flux of Uranus can be also seen in Figure 9, as reported by the different observations collected over several decades, and therefore the here reported "reference" spectrum of Uranus should be understood as a representative measure, that slowly evolves overtime due to differences in the sampled regions of the planet.

**R2017 Observations:** As previously presented (see the Jupiter section for more detail), the spectrum for Uranus from Roberge et al. (2017) was created using data from Karkoschka 1998, the SpeX instrument on the NASA Infrared Telescope Facility (Rayner et al. 2003), and model data (Burrows et al. 2004). Here we include the portion of their data set that corresponds to the observations made with the SpeX instrument, spanning 1.002 to 2.497 μm, and scaled in their work to align with the continuum from Karkoschka 1998 (K1998).

**I2022 Observations:** Reflected light observations from HST Space Telescope Imaging Spectrograph (STIS) and the NASA Infrared Telescope Facility (IRTF) are adopted and reduced in the analysis by Irwin et al. 2022. The HST data were obtained on August 19, 2002, and published in Karkoschka and Tomasko 2009. Results are published as I/F values recorded from the central meridian to the edge of the planet. The spectrum spans 0.302-0.999 μm at a resolution of 1 nm, sampled every 0.4 nm. The flux values are divided by the solar flux spectrum by Colina et al. 1996 in the analysis from Karkoschka and Tomasko 2009. The HST data set is scaled to align with the reference continuum (K1998) and multiplied by an empirical scale factor of 0.95 in Figure 9B. Additional observations are included from IRTF SpeX on May 18, 2000. The data were recorded with a long-slit spectrometer aligned on the disk center and the fluxes integrated along the central meridian. The IRTF spectrum spans a wavelength range of 1.003- 2.473 μm. The data collected from HST were smoothed to spectral resolution of 2 nm of the IRTF SpeX instrument, with 1 nm sampling to allow for a direct comparison between the two sources. The IRTF observations are included in the composite spectrum for Uranus.

## 11 Neptune

### 11.1 Neptune Introduction

Like Uranus, Neptune is classified as an "ice giant" and orbits at approximately 30 astronomical units from the sun, making it the most distant planet in our solar system. Due to the similarity in mass between Uranus and Neptune (14.5 and 17.1 Earth's mass respectively) and density (1.27 and 1.64 g/cm$^3$) the common characteristics suggest they form their own class of gas giants (Reinhardt et al. 2019). Voyager 2 is the only spacecraft to observe the planet in close proximity, and it uncovered countless insights into the planetary environment. Similar to its study of Uranus, in August 1989 Voyager 2 collected radio tracking data during an occultation of Neptune to study thermal structure and composition of the troposphere and stratosphere. Storm systems were discovered to be similar to those found on Jupiter, a thin methane cloud at about 1.5 bar pressure level was discovered, and optically thick clouds below 3 bars were observed. Additionally, two narrow rings, two broad rings, and six new moons were observed on Neptune and new surface and atmospheric features were observed on Triton (Stone and Miner 1989). The results from this study allowed for detailed investigations of the vertical structure of Neptune's atmosphere including the ionosphere, the thermal structure and composition of its troposphere and stratosphere, and zonal winds below 1.7 bars in the troposphere (Lindal et al. 1990, Lindal 1992). Additionally, the spacecraft collected over 9000 images of the planet over a 6 month period surrounding its closest approach, revealing highly detailed cloud top features (Smith et al. 1989). Currently there are no planned missions to Neptune, however, several mission concepts have been proposed. In this work we consider reflectance spectra of Neptune from a combination of ground and space-based instruments.

### 11.2 Neptune Data

**Neptune Composite Spectrum:** Using a variety of publications and data from both ground- and space-based instruments, we have constructed a composite geometric albedo spectrum of Neptune spanning 0.221 to 2.496 μm. In this section, the more recent Irwin et al. 2022 (I2022) data serve as the main reference spectrum that each of the other sources were scaled to. In the composite spectrum, data from the Hubble Space Telescope, published in Courtin 1999 are included between 0.221 and 0.320 μm. Observations from the European Southern Observatory are included from Karkoschka1998, between 0.320 and 1.000 μm. Additionally, observations from the Infrared Telescope Facility are included, published in Irwin et al. 2022 between 1.004 and 2.496 μm. The composite spectrum contains scaled versions of the original data, using the factors described below and in Table 9. Similar to Uranus, Schume et al. 2016 reports a progressive brightening of Neptune over the last decades, and several percent since the measurements reported in Karkoschka 1998. These variations, together with localized regional variability, are primarily related to changes in high altitude clouds, which evolve over long timescales. As shown in Figure 10, these brightening in the optical can be seen panel A, with the more recent measurements displaying higher values in the 0.4-0.6 μm spectral range. As such, we use the most recent measurements by Irwin et al. 2022 (I2022) as photometric reference in this work.

**FL1979 Observations:** Spectral observations from Fink and Larson 1979 were collected using the Kitt Peak National Observatory's 4 m telescope and rapid scanning Michelson interferometer. Neptune was observed on 3 observing nights between February 28th and March 2nd, 1975. The data span 0.939 to 2.450 μm at a spectral resolution of 46 $cm^{-1}$. The fluxes in their work were divided by the continuum of a reference star and the results were then corrected to align with published results from Wamsteker 1973 to produce the geometric albedo spectrum. The data were multiplied by an empirical scaling factor of 0.7 in Figure 10B to better align with the reference spectrum Irwin et al. 2022 (see below I2022).

**N1984 Observations:** Neptune was observed with the McDonald Observatory's 2.1-m telescope and the ES-2 spectrograph on May 17-18, 1981, and results are included from Neff et al. 1984. The planet was observed at a phase angle of approximately 0˚ on each night, and geometric albedo was computed from monochromatic flux ratios by using equation 1 in Neff et al. 1984. The values were published in their work from 0.350-1.051 μm at a spectral resolution of approximately 0.7 nm. The data were scaled to align with the reference spectrum Irwin et al. 2022 (see below I2022) , and multiplied by an empirical scale factor of 1.25 in Figure 10B.

**W1986 Observations:** Observations in Wagener et al. 1986 were collected using the large aperture of the Long Wavelength Prime (LWP) camera on the International Ultraviolet Explorer (IUE) satellite. Neptune was observed several times between October 2nd and 4th, 1985 and data were calibrated and combined to produce geometric albedo values. Three different geometric albedo curves were produced in their work by dividing the planetary fluxes by the scaled and extinction corrected stellar fluxes of 2 solar analogs (16 Cyg A and B) and the 1980 solar spectrum from (Mount and Rottman 1981, Mount and Rottman 1983). The results of the three curves were averaged to produce the final geometric albedo curve included in this work, spanning from 0.205 to 0.334 μm. The standard deviation of the combined data set is described in their work.

**K1998 Observations:** Neptune was observed using the European Southern Observatory's 1.52 m telescope, and results were published in Karkoschka et al.,1998. A full-disk albedo spectrum, ranging from 0.30 to 1.05 μm, was produced using data collected between July 6-10, 1995. The data were recorded with a spectral resolution of 0.4 nm between 0.300-0.995 μm and 1 nm between 0.995-1.050 μm. Neptune was observed at a phase angle of 0.3° on July 7, 1995, so the phase difference from zero was negligible. The publication presents the full-disk albedo as I/F data averaged over the planet's disk, taken at a specific phase angle, while geometric albedo refers to the same value at phase zero. Data from this source are scaled by 1.1 to align with the more recent reference spectrum.

**C1999 Observations:** Like Uranus, reflected light spectra were also recorded with the Hubble Space Telescope's Faint Object Spectrograph (FOS) in Courtin 1999 on June 22nd and August 19th, 1992, for Neptune. Data were recorded as part of the GTO Program 1290-GTO/ OS-86G. Calibration of the data is similar as that reported for Uranus in the previous section and scaled by 1.1 to align with the more recent reference spectrum of Neptune by Irwin et al. 2022 (see below I2022).

**R2017 Observations:** As previously presented (see the Jupiter section for more detail), the spectrum for Neptune from Roberge et al. (2017) was created using data from Karkoschka 1998, the SpeX instrument on the NASA Infrared Telescope Facility (Rayner et al. 2003), and model data (Burrows et al. 2004). Here we include the portion of their data set that corresponds to the observations made with the SpeX instrument, spanning 1.00 to 2.50 μm, and scaled in their work to align with the continuum by Irwin et al. 2022.

**I2022 Observations (photometric reference):** As previously discussed for Uranus, the spectra in Irwin et al. 2022 contain a combination of NASA-IRTF and HST datasets, spanning the 0.3 to 2.5 μm spectral range. The HST are included in the composite reference spectrum for Neptune between 0.3 to 1.0 μm and the IRTF data within the 1.0 to 2.5 μm spectral range.

## 12 Discussion

### 12.1 Color Index Analysis

Visible planetary magnitudes and color indexes have historically been considered as a preliminary tool to broadly classify solar system objects. Reflectance spectral slopes, or broad band photometry color-color diagrams at the visible wavelengths were proposed early on as a means of distinguishing rocky versus gas giant planets (Traub 2003a). Scientists theorized that since the planetary flux is dependent on the albedo and radius of the planet, while the mass depends on the radius and mean density, then the link between these characteristics may provide a strong foundation for grouping based on planetary types. By taking the difference of the blue and green magnitudes ($m_B$-$m_G$) and plotting it against the difference in the green and red magnitudes ($m_G$-$m_R$), Traub et al., showed that the planets fall into 3 main categories (Traub 2003b). The rocky planets (including Mercury, the Moon, and Mars) cluster in the red-red region of the albedo colors plot while the ice giants (Uranus and Neptune) cluster in the green-blue region. Additionally, cloudy outer planets (Jupiter, Saturn, and Titan) cluster together in center of the diagram, while Earth and Venus tend to separate from all other planets.

A different study mapped planets and smaller solar system bodies based on three distinct compositional groups as identified by color-color maps in the near-IR: gas planets, soil planets, and ice planets. This classification relied on low resolution, broad band spectroscopy in the visible and near infrared (Lundock et al. 2009). They plotted color-color diagram of J (1.25 μm) – K (2.15 μm) versus Rc (0.66 μm) – J (1.25 μm) magnitudes and found the method to be effective for sorting the planets and the moons into the three categories, however, this methodology failed in categorizing Venus and Ganymede in the correct physical classification since they would have been misinterpreted them at first glance, indicating that this method should be only used as an first estimate of planetary types.

Another method of color-color classification was presented by Crow et al., in which the ratio of reflectance intensities at 850 nm/550 nm was plotted against those at 350 nm/550 nm. In this color-color plot , it was found that the planets separated into four main categories (Crow et al. 2011). The airless bodies were clustering on the right side of the diagram, the intermediate cloudy atmospheres grouped near unity, while the strong NIR absorbing atmospheres fell on the left of the plot. The planets with intensities ratios larger than 1 were interpreted as indicative of strong Raleigh scattering, and the only planet existing within the upper right quadrant was Earth, due to its Raleigh scattering in its intermediate cloudy atmosphere.

These studies show a few examples of techniques that scientists have previously adopted to classify planetary atmosphere and characterize them, in the broader context of exoplanetary research. However, these broad band photometry techniques and color-color diagram plots showed that they could be misleading in the interpretation and physical classification of planets. This is why, in our first portion of our study we collected and created a database of spectra of each planet of the solar system. Spectroscopy is necessary to characterize planetary environments, especially in the framework of identifying habitable exoplanets around other stars. It is nevertheless important to notice that spectroscopy is not always available, so colors can still be an informative tool to start a first characterization of faint exoplanets.

***Planets RGB colorimetry:*** In this section we use the visible portion of our collected reference planetary spectra from 0.35 to 0.7 um, to create the true RGB color representation of each of the solar system planets. To do this, we used a similar approach as described in Irwin et al. (Irwin et al., 2024), of reconstructing the true colors of a planetary body by adopting RGB filters properly adjusted and designed to replicate the cone responsivities of the average human eye (Reye, Geye, Beye). While Irwin et al. specifically concentrated on producing true RGB colors for Uranus and Neptune, adding realism to their images that had previously not existed, in our work we produce true RGB colors of all the planets in the solar system, and Titan. To do this, we first digitized the eye red, green and blue filters from Irwin et al. (RGB hereafter), then we took our collection of disc-averaged planetary reflectance spectra, and we transformed them from reflectance, $A_p$, to Flux (in units of $W/m^2/\mu m$), with the following equation:

$$F_p = A_p * \frac{F_s}{R_h^2} * \frac{\pi D_p^2}{\Delta^2} \; (W/m^2/\mu m)] \tag{3}$$

where $F_s$ (in units of W / $m^2$/μm) is the flux of the Sun at 1 (au), $R_h$ (au) is the heliocentric distance of the planet, $D_p$ (km) is the planet radius, Δ (km) is the distance to the observer, which in this case we are assuming to be placed at the position of Sun. Our next step was to extract the flux integrated over each of the RGB bandpasses, previously normalized to their bandpass area, as following:

$$F_p^{Filter} = \int F_p(\lambda)\ \ \eta^{Filter}(\lambda)\ \ d\lambda \tag{4}$$

where $\eta^{filter}$ is the filter spectral response and λ is wavelength. In Table 10, we report the values of heliocentric distances, the planet diameters and the computed RGB values, normalized to the maximum, while in Figure 11 we show the true colorimetry of the solar system.

***Planets absolute magnitude H and color index B-V:*** The analysis in this portion follows the same approach as Mallama et al. 2017, and we compile an updated list of reference B and V magnitudes in the Johnson-Cousins broad-band photometric system (Johnson et al., 1966; Cousins, 1976a, Cousins 1976b). To do this, we first downloaded the Bessell B and V optical glass filter bandpasses (Bessel 1990), which reproduce reasonably well the classic Johnson-Cousins passbands (http://spiff.rit.edu/classes/phys440/lectures/filters/filters.html). Then we determined the contributions of the planetary flux integrated over the B and V Johnson filters, previously normalized to their band area. The B and V magnitudes of the planets were then determined adopting the following equation:

$$m = -2.5\ log\ F\ -\ m^0 \tag{5}$$

where $m^0$ is the magnitude of Vega ( https://www.astronomy.ohio-state.edu/martini.10/usefuldata.html ). The B-V color index is determined by simply taking the difference between the two magnitude values.

$$B - V = \ (m_B -\ m_V) \tag{6}$$

The B-V color indices are validated with results from Mallama et al. 2017. In their work they used a collection of observations (see references therein) and synthesize magnitudes for each of the solar system planets. Their study includes photometric values across the standard Johnson Cousin (U,B,V,R,I) for each of the solar system planets. The results were used to compute the color index (B-V) and the geometric albedo values were compared with the results from our green and Johnson V filters in Table 10. Lastly, we computed the absolute visual magnitude $H_v$ for the planets, which is defined as the apparent magnitude of the object illuminated by the solar light flux at 1 (au) and observed from the distance of 1 (au) and at zero phase angle (Pravec and Harris 2007), and follows the equation below:

$$H_v = 5\frac{1329\ (km)}{D(km)\sqrt{Ageo}} \qquad (7)$$

where ($A_{geo}$) is the geometric albedo, ($D$) the diameter for each planet, and the scale factor of 1329 (km) is linked to the apparent magnitude of the Sun (Bowell et al. 1989, Pravec and Harris 2007). In Table 10 we report the apparent magnitude $H_{vis}$, and compare it to Mallama et al., 2017, the geometric albedo $A_{geo}$ and the B-V color index for each planet, as well as a comparison to previous works. In Figure 12 we instead show the visual portion of the planetary spectra between 0.35 and 0.7 μm organized by the B-V color index values, with the bluer spectra to the bottom and the redder spectra to the top of the plot. The digitized filters bandpasses are displayed alongside our results describing the regions where flux was determined for each of the RGB colorimetry study and for the B-V color index analysis. The figure illustrates how the changing of the slope in this portion of the spectrum changes our color perception of each planet, where positive slopes are indicative of a redder planet and negative slopes correspond to bluer planets. Extracting this information directly from spectroscopic observations is incredibly valuable as it allows us to produce the visuals RGB colorimetry shown in Figure 11 with high accuracy.

The importance of investigating planets' colors resides in their connection and support to exoplanetary research, especially when spectroscopy is not available. The identification and quantitative characterization of astrobiologically relevant atmospheric features, such as water vapor, ozone, or biosignature gases is robustly done via spectroscopy. However, spectroscopy is less common and not as straightforward as color photometry, which can be easily obtained using ground-based telescopes.

Colorimetry and B-V index classifications may be used, with caution, to assist in categorizing solar system planets and exoplanets (e.g., hot Jupiters, super-Earths, etc.), by distinguishing the different types of planets, based on their optical properties. These optical properties contain information about surface materials and atmospheric properties, since minerals, ices, and atmospheric compounds have distinct spectral signatures. Therefore, this methodology can be used as a preliminary means of distinguishing between rocky, icy, and gas-rich bodies and support the identification of potentially habitable worlds. It is important to note that this method has some intrinsic degeneracies. For example, similar colors may result from different atmospheric compositions and colors can vary depending on the observed planetary phase. Additionally, Rayleigh scattering makes atmospheres appear bluer, while cloud coverage can mask surface and atmospheric features. Ultimately, broad-band photometry in the optical can provide a cost-effective way to gather initial characterization data, however more detailed spectroscopic follow-up is then required.

### 12.2 Reflectance Spectra Interpretation for Exoplanet Analogs

In this section, we consider how the I/F spectra that were created in the prior sections may be implemented to simulate observations of our solar system from afar and nearby planetary systems using their solar system analogs. Anterior to our work, Roberge et al. (2017) simulated spectral flux densities for each of the solar system planets (excluding Mercury) as seen at quadrature from 10 pc away (their Figure 4). Similar to our work, they incorporate Karkoschka (1998) spectra to calibrate the observations for Jupiter, Saturn, Uranus, and Neptune. The giant planets in their study combine Karkoschka (1998) and data from the SpeX instrument on the NASA Infrared Telescope Facility (Rayner et al. 2009). However, for Venus, Earth, and Mars they used simulations from the Spectral Mapping Atmospheric Radiative Transfer (SMART) model (Meadows & Crisp 1996; Crisp 1997) from the Virtual Planet Laboratory. In their work, they also incorporate exozodiacal dust scattering models.

All the spectra in our study come from calibrated observational data for each of the solar system planets and Titan and are displayed here as they would be seen from 10 parsecs away. We use the solar system planetary spectra to represent analog planets in nearby multi-planet systems around different star types. We selected HD 219134 (K3V, Keenan and McNeil 1989) as our nearby multi-planet K dwarf system and Proxima Centauri (M5.5V, Bessel 1991) as our nearby multi-planet M dwarf system. The objective is to compare how the same planetary spectra on analogous planets from another system would have different levels of detectability depending on the planetary system and host star characteristics. Both systems are within 10 parsecs of our solar system making them potentially strong candidates for future direct imaging missions. Our first step was to identify which planets within these two systems are analogous to our solar system planets, to associate to them their reflectance spectra. The habitable zone (HZ) boundaries for each exoplanetary system are defined using the conservative limits from Kopparapu et al. 2014.

Proxima Centauri is the closest star to our Sun, sitting at a distance of only 1.30 parsecs (Demasso et al. 2020). It is a red dwarf star that is much smaller, cooler, and denser than our sun, containing three known planets, all were detected using the radial velocity method. Proxima Centauri b was detected in 2016 (Anglada-Escudé et al. 2016) and is the only confirmed planet in the system, similar in size to Earth (~1.07 $M_{Earth}$, Artigau et al. 2022) and orbiting within the habitable zone (0.049 au, Faria et al. 2022), where liquid water could potentially exist on the surface. In absence of known transit, the radius is not well constrained for this planet. We have assumed Earth density of 5.51 g/cm$^3$ to derive a radius of 1.02 $R_{Earth}$. Proxima Centauri c and d are both candidate planets (due to less extensive validation through observations) detected in 2020 (Demasso et al. 2020) and 2022 (Faria et al. 2022) respectively. Planet c is believed to have a minimum mass of approximately 5.8 times the Earth (Demasso et al. 2020), and orbits at about 1.48 au (Demasso et al. 2020), far beyond the habitable zone boundary for such red star. Planet c reflectance was therefore simulated by adopting our reflectance spectrum for Mars. Similar to planet b, the radius is calculated in this work using its mass from Demasso et al. (2020) and the density of the analog planet, Mars. Finally, planet d orbits closest to the star with a semimajor axis of 0.029 au (Faria et al. 2022), with a minimum mass of 0.26 $M_{Earth}$ (Faria et al. 2022); it is represented using our reflectance spectrum for Mercury.

HD 219134 is a main sequence star classified as K3V sitting at 6.53 parsecs (Keivan et al. 2019) and hosting the next planetary system considered in our analysis. This star is a bit cooler and smaller than our Sun with a black body temperature of 4817 K (Rosenthal et al. 2021) and only about a quarter of its luminosity (Li et al. 2025). It hosts six confirmed exoplanets, all discovered using the radial velocity method in 2015 (Vogt et al. 2015). The HD 219134 system has a habitable zone between 0.372 and 0.784 AU. The innermost planets in the system, b and c, are the only ones known to be transiting, they have semimajor axes of 0.039 (Gillon et al. 2017) and 0.065 au (Gillon et al. 2017) respectively, both sitting outside of the inner habitable zone boundary. Planet b has a mass of 4.59 $M_{Earth}$ (Li et al. 2025) while planet c has a mass of 4.23 $M_{Earth}$ (Li et al. 2025) and both were represented using our Mercury reflectance spectrum. Planets d and f also sit outside of the habitable zone with semimajor axes of 0.237 au (Gillon et al. 2017) and 0.146 au (Gillon et al. 2017) respectively. Planet d has a minimum mass of 16.17 $M_{Earth}$ (Gillon et al. 2017) and planet f has a minimum mass of 7.3 $M_{Earth}$ (Gillon et al. 2017); both are simulated with our Venus reflectance spectrum. Planet g sits inside the inner edge of the habitable zone, with a semimajor axis of 0.375 AU. Since the radius is not well constrained for planet g, we computed it using the planet's mass (~10.8 $M_{Earth}$; Vogt et al. 2015) and the density of Neptune (the analog planet) at 3.31 $R_{Earth}$. Planet h sits beyond the habitable zone, with a semimajor axis of 3.11 au (Vogt et al. 2015) and a mass of ~76 $M_{Earth}$. We assumed it to be a Jupiter analog and using Jupiter's density of 1.33 g/cm$^3$ we obtained a 6.8 Earth radius in the cold Jovian planet category following Kopparapu et al. (2018) classification.

To compare how our solar system planetary spectra would look like from a distant system, our approach consists in transposing the solar system planets' observed data to the exoplanets in those systems. The objective is not to self-consistently predict the reflectance spectra of these exoplanets but to show how our solar system planetary spectra could look like under different system configurations. For the solar system as seen from 10 pc away, we first converted the *I/F* reflectance spectra to absolute flux density units (Jy). This conversion consists in multiplying the *I/F* spectrum by the spectrum of the Sun (*F*) as seen from the top of the atmosphere (TOA) of the planet to obtain *I* (in units of W/m$^2$/sr/μm). Then, a disk-integration was performed accounting for the planet's disk surface area and the system distance of 10 pc, considering phase 90 (quadrature). Scaling to quadrature from our geometric albedo reference was done by adopting the appropriate phase integral for each case. The total planetary signal I was computed by adding this computed reflected signal to the thermal flux. Finally, the quantity *I* was converted from $W/m^2/\mu m$ to an absolute flux unit, Jansky (Jy).

For HD 219134 and Proxima Cen, the star spectral energy distribution (SED), the planet-star distances, and planetary diameters are different from our solar system planets. To transpose our solar system planet's spectra to the closest analogue planets in those systems, we first used PSG to simulate an idealized flat disk spectra for those exoplanets. Those flat disks consist in simulating the planet's reflected flux (Jy) with their true orbital distance, size, instellation (including the SED), and distance of the system, at phase 0 and assuming a perfectly reflecting and emitting surface (albedo=emissivity=1). The equilibrium temperature of each planet was determined considering a Bond albedo as the corresponding of its solar system analogue. Effectively, such a flat disk spectrum is the geometric albedo of a planetary disk perfectly reflecting the instellation *F* that the exoplanet should receive (along with the self-emission of the planet) to have the same *I/F* as its solar system analogue. The reflected flux of these flat disks was then multiplied by the representative spectrum of each planet while also considering a phase correction to obtain the reflected flux in Jy at the true distance of the system. The total planetary signal I was computed by adding this computed reflected signal to the thermal flux, and then I/F was computed. The parameters used to convert the reflectance spectra to spectral flux densities are described in Table 11.

The results of such analysis are presented in Figure 13. First, on the upper panel we can notice large differences in the SED, both in term of location of the peak and of amount and type of spectral lines with the later type stars displaying a stronger non-black body SED. We can see without surprise that the Solar System reflected light spectra is dominated by Jupiter, and then Venus. The lowest spectral flux density belongs to the confine of the solar system (Uranus,

Neptune and Titan). For the K-dwarf system HD219134, we can see that the Mercury analogs planets b and c strongly dominate. Indeed, those two transiting planets are the closest to the star (0.039 and 0.065 au, respectively), followed by the Venus analogs planets d and f (at 0.237 and 0.146 AU) and the Neptune analog (planet g at 0.375 AU) and the Jupiter analog (planet h at 3.11 AU). Contrary to our Solar System dominated by Jupiter reflectance, planet h with its 0.61 Jupiter radius and 79 % of Jupiter insolation appears relatively dimmer than the other planets in the system. For Proxima centauri, planet d (Mercury analog) spectral flux density dominates over planet b (Earth analog) and planet c (Mars analog) due to its higher insolation. Planet b and c spectra look very similar in the visible range but start to diverge in the near infrared. For the three planets analog, the spectral flux density at the 1.30 parsec distance of the Proxima cen system is relatively close to what we expect from those planets in our solar system as seen from 10 pc away, the difference in the star bolometric luminosity and distance of the systems being the major factors playing in that coincidence. Overall, the figure reveals that depending on both the instellation and the size of the planet, the relative reflectivity of a given planet's atmosphere with respect to the other planets in the system would appear very different from a system to another.

### 12.3 Future Direct Imaging Mission Analysis

Using the reference spectra of diverse planetary bodies in our solar system and scaled to distant systems as reported in the previous section, we can now explore how these objects would be seen if observed with advanced space observatories. Direct imaging via coronagraphy is currently the main technique under consideration for the study of potentially habitable exoplanets. Coronagraphy involves blocking the strong signal from the central star and integrating the signature of the nearby orbiting planet. The challenge is that habitable planets like

Earth are extremely small relative to the host-star, requiring extraordinary contrasts and occultation of the central star, in the order of $10^{-10}$ (i.e. 60 versus ~$10^{-8}$ Jy in our Figure 13). In order to address this challenge, NASA is developing the Habitable Worlds Observatory (HWO), a space telescope designed to directly image Earth-like exoplanets around nearby stars and assess their potential habitability. Planned for launch in the 2040s, HWO builds on technologies from missions like *Hubble Space Telescope*, *James Webb Space Telescope*, and *Roman Space Telescope*, but with a primary focus on detecting biosignatures on distant worlds. A key feature of HWO is its coronagraph, a highly advanced instrument that blocks out the overwhelming light of a host star to reveal the much fainter planets orbiting around it. The coronagraph uses deformable mirrors and precision optics to suppress starlight by a factor of at least 1 part in 10 billion, enabling the telescope to capture direct images of rocky planets in the habitable zones of Sun-like stars—something no current telescope can do. This capability would be critical for analyzing planetary atmospheres and searching for gases like oxygen ($O_2$), water ($H_2O$) or methane ($CH_4$) that may hint at life.

Several designs are being considered for the coronagraph, each exploring a different trade on contrast, throughput and overall performance. Designs include the Shaped Pupil Coronagraph (SPC) and Hybrid Lyot Coronagraph (HLC) from the Roman Space Telescope, as well as the Vector Vortex Coronagraph (VVC) and Apodized Pupil Lyot Coronagraph (APLC) from the LUVOIR and HabEx concepts. In order to test the impact on the design parameters of the coronagraph, we pick one of the preferred designs considered for HWO, the Deformable Mirror-Apodized Vortex Coronagraph (DMAVC). When summarizing the performance of a coronagraph, four key performance metrics—inner working angle (IWA), outer working angle (OWA), core throughput, and contrast—collectively determine the system's ability to detect and characterize exoplanets. The inner working angle (IWA) refers to the angular separation at which the contrast delivered by the coronagraph starts to degrade substantially. This is a critical parameter because it determines how close to a star the instrument can observe, which is essential for finding planets in tight orbits, such as Earth-like planets within habitable zones. The IWA is typically proportional to the wavelength of light divided by the telescope aperture diameter ($\lambda/D$), meaning that larger telescopes or shorter wavelengths enable observations of planets at smaller separations. The latest design of the DMAVC for HWO delivers an IWA of ~2.5 $\lambda/D$. The outer working angle (OWA) defines the largest angular separation from the star where the coronagraph still maintains effective suppression of starlight. It marks the outer limit of the instrument's field of view for planet detection, allowing observations of more distant planetary companions. The OWA is influenced by the coronagraph's optical design and the size of the focal plane mask or detector array, and it is estimated to be ~30 $\lambda/D$ for the DMAVC design.

Core throughput is the fraction of a planet's light that passes through the coronagraph and is successfully recorded by the detector, relative to what would be detected in the absence of any starlight suppression system. Higher core throughput indicates that more of the exoplanet's signal is preserved, which improves sensitivity and reduces the required exposure time. Throughput is affected by the optical efficiency of components such as masks, apodizers, and wavefront control elements, and it's relatively high (~0.45) at 20 $\lambda/D$ for the DMAVC design. Finally, contrast refers to the ratio of the brightness of the star to the faintest planet signal that the coronagraph can detect at a given angular separation. It measures how well the coronagraph suppresses starlight and is crucial for distinguishing faint exoplanets from the overwhelming glare of their host stars. The DMAVC design is defined based on the required contrast to detect Earth-like planet detection around Sun-like stars, and it is $10^{-10}$ within the 2-30 λ/D range.

Using the reported contrast and core throughput performance curves from Belikov et al. (2023) for the DMAVC coronagraph design, we simulated planetary fluxes adopting a telescope with a 7.14-meter circumscribed diameter for HWO. Planetary reflectance was modeled for Earth- and Jupiter-like analogs using geometric albedos previously described in Sections 4 and 6. All planets were assumed to be observed at quadrature (phase angle of 90°), and a phase correction factor was applied to adjust the geometric albedo accordingly. The integrated planetary fluxes decrease with the square of the distance, reaching marginal levels (<1 nJy) for Earth-like planets at distances beyond 20 parsecs. As illustrated in Figure 14, the influence of the inner working angle (IWA) becomes increasingly significant at longer wavelengths, effectively obscuring key spectral features beyond 1.5 μm for Earth analogs located farther than 10 parsecs. Likewise, the outer working angle (OWA) imposes constraints for giant planets at large orbital radii: for Jupiter analogs within 10 parsecs, a substantial portion of the reflected optical spectrum lies outside the OWA, limiting detectability. These geometric and instrumental constraints underscore the dependence of system characterization capabilities on both stellar distance and the specific coronagraphic architecture. Consequently, the ability to detect and analyze biosignature-bearing molecules (e.g., $O_2$, $O_3$, $H_2O$, $CO_2$, $CH_4$) and atmospheric constituents such as aerosols and cloud layers (e.g., photochemical haze, water ice clouds) may be significantly restricted by the combined effects of IWA, OWA, and system sensitivity.

## 13. Conclusions

As new mission concepts aim to directly image distant planetary systems and characterize potentially habitable Earth-like exoplanets, it is essential to aggregate and reconcile decades of astronomical studies targeting the planets that we know best. In this work we have combined the existing resources to provide a comprehensive framework that will aid in our interpretation of future exoplanet atmospheric detections. Sources were compiled from an extensive literature search of observations from ground and space-based observatories, with geometrical and calibration factors determined to correct for differences in observational geometries, instrumental variability, and reported intensity units. This inter-calibration and aggregation of the datasets allowed us to properly compare the reported reflectances and determine distinctive and unique consistent spectroscopic features for each planet.

The goal of this work was to provide the spectral library as a resource that can be downloaded and adopted as a reference for atmospheric modeling and in the context of simulating results applicable to upcoming direct imaging missions, such as the Habitable Worlds Observatory. The discussion demonstrated how the inner and outer working angles impose limitations on our ability to detect certain atmospheric absorption features as a function of system distance in parsecs. In addition, we explored how nine different exoplanets from the Proxima Centauri and HD 219134 systems may be simulated as solar system analogs by adopting the reference spectra from our library. The solar system planets and Titan are also simulated as they would be directly imaged at a distance of 10 parsecs. Finally, the visible light portion of our results were used to generate accurate color reconstructions of the disk integrated view of each planet in our study, compute B-V color indices, and determine the visible geometric albedo values for validation with previous work, which can be used to calibrate future and past photometric astronomical datasets.

## Acknowledgments

The team acknowledges support from the GSFC Sellers Exoplanet Environments Collaboration (SEEC) and the ExoSpec work package, which are funded by NASA's Science Mission Directorate (SMD) Internal Scientist Funding Model (ISFM). We thank Prof. Tyler Robinson and the anonymous reviewers for their constructive comments and insightful suggestions, which significantly improved the clarity and quality of this manuscript.

| **Table 1: Mercury** | | | | | | | | | |
|---|---|---|---|---|---|---|---|---|---|
| **Source** | **Ref.** | **Mission/ Instrument** | **Date** | **Original units** | **Wavelengths (μm)** | **Phase Ang. (°)** | **Geometry** | **Geometric Scaler** | **Empirical Scaler** |
| MC1979 | McCord and Clark 1979 | 2.24-m telescope on Mauna Kea, Hawaii | 04/21/1976 | Reflectance (Mercury/Sun) | 0.650-2.200 | 77.4 | Disk Integrated | 2.102 | 0.660 |
| D2010 | Domingue et al. 2010 (Fig. 8) | MESSENGER/ MDIS, MASCS, and VIRS | 12/15/2008 | I/F | 0.308-1.460 | 74 | Disk Integrated | 1.986 | 4.500 |
| I2014 | Izenberg et al. 2014 (Fig. 5a) | MESSENGER/ MASCS | n/a | Reflectance | 0.311-1.421 | n/a | Averaged Local measurements | 1 | 2.000 |
| M2017 | Mallama et al. 2017 (Table 7) | SOHO, 2000-mm focal length Schmidt-Cassegrain | Various nights between 06/06/1999 - 07/08/2000 | Geometric Albedo | 0.360-0.900 | 0 | Disk integrated | 1 | 1 |
| K2018 | Klima et al. 2018 (Fig. 2a) | MESSENGER/ MDIS | 03/18/2011 ,04/30/2015 | Reflectance | 0.428-0.999 (HRM), 0.427-0.998 (LRM) | n/a | Local measurements: Northern polar region (HRM), Equator- South polar region (LRM) | 1 | 2.000 (HRM), 3.200 (LRM) |

**Table 1**: Parameters describing each source used in the analysis of Mercury are reported. In order of appearance, the label used to describe the data (column 1), literature reference (2), mission or instruments used (3), date of the observations (4), units from the original publication (5), wavelength range in μm (6), and phase angle in degrees at the time of the observations (7, also plotted in Figure 2A). Columns 8-10 report the geometry of the observation, the geometric scaling factor applied to normalize to a standard full-disk geometry (plotted in Figure 2B), and the empirical scaling factor used to match the geometric albedo data from the Mallama et al. 2017 photometric reference spectrum (plotted in Figure 2C).

**Table 2: Venus**

| **Source** | **Ref.** | **Mission/ Instrument** | **Date** | **Original units** | **Wavelengths (μm)** | **Phase Ang. (°)** | **Geometry** | **Geometric Scaler** | **Empirical Scaler** |
|---|---|---|---|---|---|---|---|---|---|
| I1968 | Irvine 1968 (Tab 2) | Harvard College Observatory | 1963-1965 | Spherical (bond) Albedo | 0.315-1.06 | n/a | not reported | 0.66 | 1.10 |
| W1972 | Wallace et al. 1972 (Fig 2) | OAO-2 | February 16, 1969 | Albedo | 0.215-0.300 | 103 | not reported | 4.67 | 0.85 |
| B1975 | Barker et al. 1975 (Fig 6) | McDonald Observatory, 2.7 m | May-July 1974 | Spherical Albedo | 0.308-0.592 | n/a | Disk integrated | 0.66 | 1.11 |
| M1983 | Moroz 1983 (Tab. 3) | Various instruments and observations | Multiple dates | Single Scattering Albedo | 0.320-3.800 | Multiple rescaled to 0 | Multiple | 0.66 | 1.23 |
| M2017 | Mallama et al. 2017 (Tab. 7) | SOHO/LASCO & Schmidt-Cassegrain Telescope | Multiple Years 1999 to 2004 | Geometric Albedo | 0.360-0.798 | 0 | Full-disk | 1.00 | 1.00 |
| PH2018 | Perez-Hoyos et al. 2018 (Fig8) | MESSENGER/VIRS | June 5, 2007 | I/F | 0.308-1.492 | n/a | Nadir (equatorial atmosphere) | 1.00 | 1.13 |
| V2019 | Vlasov et al. 2019 (Fig 1) | VEX/SPICAV | 8/ 9/ 2006 | Albedo | 0.202-0.308 | n/a | Combined nadir observations | 1 | 1 |
| L2021 | Lee et al. 2021 (Fig1) | Venus Climate Orbiter /UVI | 12/ 2015-01/ 2019 | Albedo | 0.283,0.365 | 3-10 | Disk integrated | 1 | 1 |

**Table 2:** Parameters describing each source used in the analysis of Venus are reported. In order of appearance, the label used to describe the data (column 1), literature reference (2), mission or instruments used (3), date of the observations (4), units from the original publication (5),

wavelength range in μm (6), and phase angle in degrees at the time of the observations (7, also plotted in Figure 3A). Columns 8-10 report the geometry of the observation, the geometric scaling factor applied to normalize to a standard full-disk geometry (plotted in Figure 3B), and the empirical scaling factor used to match the geometric albedo data from the Mallama et al. 2017 photometric reference spectrum (plotted in Figure 3C).

| **Table 3: Earth** | | | | | | | | | |
|---|---|---|---|---|---|---|---|---|---|
| **Source** | **Ref.** | **Data Type** | **Mission/ Instrument** | **Date** | **Original units** | **Wavelengths (μm)** | **Phase Ang. (°)** | **Geometry** | **Geometric Scaler** |
| L2011 | Livengood et al. 2011 (Fig. 4) | Observations | Deep Impact spacecraft, EPOXI | 03/18/2008 | Reflectivity | 0.369-0.965, 1.055-2.675 | 57.7 | Disk-Integrated | 1.582 |
| R2011 | Robinson et al. 2011 (Fig. 2) | Model | Spectral Mapping Atmospheric Radiative Transfer (SMART) Model, Deep Impact Spacecraft High Resolution Instrument (HRI) | n/a | Reflectivity | 0.506-1.678 | 0 | Disk-integrated | 1 |
| R2017 | Roberge et al. 2017 (Fig. 2) | Model | Spectral Mapping Atmospheric Radiative Transfer (SMART) Model, Deep Impact Spacecraft High Resolution Instrument (HRI) | n/a | Geometric Albedo | 0.300-2.497 | 0 | Disk-integrated | 1 |
| D2022 | DSCOVR_EPIC_L2_COMPOSITE_01, Marshak et al. 2018 as | Observations | Deep Space Climate Observatory (DSCOVR) | 06/21/2022 | Albedo | 0.318-0.780 | 0 | Disk-integrated | 1 |

| | reduced in K2024 | | | | | | | | |
|---|---|---|---|---|---|---|---|---|---|
| K2024 | Kofman et al. 2024 | Model | PSG/ GlobES, MODIS, MERRA-2, DSCOVR | 06/21/2022 | Albedo | 0.3-1 | 0 | Disk-integrated | 1 |

**Table 3:** Parameters describing each source used in the analysis of Earth are reported. In order of appearance, the label used to describe the data (column 1), literature reference (2), mission or instruments used (3), date of the observations (4), units from the original publication (5), wavelength range in μm (6), and phase angle in degrees at the time of the observations (7, also plotted in Figure 4A). Columns 8 and 9 respectively report the geometry of the observation and the geometric scaling factor applied to normalize to a standard full-disk geometry (plotted in Figure 4B).

| Table 4: Mars | | | | | | | | | |
|---|---|---|---|---|---|---|---|---|---|
| **Source** | **Ref.** | **Mission/ Instrument** | **Date** | **Origin al units** | **Wavelen gths (μm)** | **Phase Ang. (°)** | **Geomet ry** | **Geomet ric Scaler** | **Empiri cal Scaler** |
| MW19 71_A | Mccord and Westphal 1971 (Fig. 6) | Cerro Tololo Inter-American Observatory, 60 inch telescope | 05/25/ 1969, 05/27/ 1969 | Geom etric albedo | 0.318-1.097 | 5 | localize d | 1 | 0.7 |
| MW19 71_SM | Mccord and Westphal 1971 (Fig. 6) | Cerro Tololo Inter-American Observatory, 60 inch telescope | 05/25/ 1969, 05/27/ 1969 | Geom etric albedo | 0.316-1.099 | 5 | localize d | 1 | 1.7 |
| W1972 | Wallace et al. 1972 (Fig. 3) | Orbiting Astronomical Observatory | 04/23/ 1969 | Geom etric Albed o | 0.204-0.365 | 26.8 | Disk integrate d | 1.107 | 1 |
| M2017 | Mallama et al. 2017 (Table 7) | 2000-mm focal length Schmidt-Cassegrain | | Geom etric Albed o | 0.360-0.900 | 0 | Disk integrate d | 1 | 1 |
| R2017 | Roberge et al. 2017 (Fig. 2) | Spectral Mapping Atmospheric Radiative Transfer (SMART) Model, Deep Impact Spacecraft High Resolution Instrument (HRI) | n/a | Geom etric Albed o | 0.300-2.497 | 0 | Disk-integrate d | 1 | 1 |
| PSG Model | Villanueva et al. 2018 | Planetary Spectrum Generator (PSG) | n/a | I/F | 0.200-2.499 | 0 | Disk integrate d | 1 | 1 |

**Table 4:** Parameters describing each source used in the analysis of Mars are reported. In order of appearance, the label used to describe the data (column 1), literature reference (2), mission or instruments used (3), date of the observations (4), units from the original publication (5), wavelength range in μm (6), and phase angle in degrees at the time of the observations (7, also plotted in Figure 5A). Columns 8-10 report the geometry of the observation, the geometric scaling factor applied to normalize to a standard full-disk geometry (plotted in Figure 5B), and the empirical scaling factor used to match the geometric albedo data from the Mallama et al. 2017 photometric reference spectrum (plotted in Figure 5C).

| Table 5: Jupiter | | | | | | | | | |
|---|---|---|---|---|---|---|---|---|---|
| **Source** | **Ref.** | **Mission / Instrument** | **Date** | **Original units** | **Wavelengths (μm)** | **Phase Ang. (°)** | **Geometry** | **Geometric Scaler** | **Empirical Scaler** |
| C1979_c | Clark and Mccord 1979 (Fig. 2) | Mauna Kea Observatory, 2.24 m | 11/04/1976 | Albedo | 0.661-2.504 | n/a | Center of Disk | 0.66 | 1 |
| C1979_l | Clark and Mccord 1979 (Fig. 3) | Mauna Kea Observatory, 2.24 m | 11/04/1976 | Albedo | 0.661-2.501 | n/a | East Limb | 1 | 1 |
| C1982 | Clarke et al. 1982 (Fig. 3) | International Ultraviolet Explorer | December 1978 - July 1980 | Geometric Albedo | 0.123-0.194 | 0 | Disk averaged | 1 | 1 |
| K1998 | Karkoschka 1998 (Fig. 3) | European Southern Observatory, 1.52 m | 7/6/1995 - 7/10/1995 | Full-disk Albedo | 0.300-1.050 | 6.8 | Disk averaged | 1.007 | 1 |
| B2003 | Betremieux et al. 2003 (Fig. 1) | HST/FOS | 06/05/1993 | I/F | 0.176-0.231 | n/a | Localized Equatorial Region | 1 | 1 |
| R2017 | Roberge et al. 2017 (Fig. 2) | IRTF/Telescope SpeX | n/a | Geometric Albedo | 1.002-2.497 | 0 | Disk integrated | 1 | 1 |
| L2018_f3 | Li et al. 2018 (Fig. 3) | Cassini VIMS | October 2000 - March 2001 | Monochromatic Geometric Albedo | 0.015-3.990 | 0 | | 1 | 0.9 |
| L2018_f7 | Li et al. 2018 (Supplementary Fig. 7) | Cassini VIMS | October 2000 - March 2001 | Full-disk Albedo | 0.359-1.018 | 10.3 | | 1.016 | 1.1 |
| G2021 | Giles et al. 2021 (Fig. 1) | Juno/UVS | August 2016 - November 2020 | Reflectance | 0.140-0.0.200 | n/a | Localized Equatorial Region | 1 | 1 |
| C2022 | Coulter et al. 2022 (Fig. 4) | Cassini/VIMS | 12/12/2000 | I/F | 0.351-5.132 | 1.66 | Disk integrated | ~1 (1.0004) | 1.186 |

**Table 5:** Parameters describing each source used in the analysis of Jupiter are reported. In order of appearance, the label used to describe the data (column 1), literature reference (2), mission or instruments used (3), date of the observations (4), units from the original publication (5), wavelength range in μm (6), and phase angle in degrees at the time of the observations (7, also plotted in Figure 6A). Columns 8-10 report the geometry of the observation, the geometric scaling factor applied to normalize to a standard full-disk geometry (plotted in Figure 6B), and the empirical scaling factor used to match the geometric albedo data from the Karkoschka 1998 photometric reference spectrum (plotted in Figure 6C).

| Table 6: Saturn | | | | | | | | | |
|---|---|---|---|---|---|---|---|---|---|
| **Source** | **Ref.** | **Mission/ Instrument** | **Date** | **Original units** | **Wavelengths (μm)** | **Phase Ang. (°)** | **Geometry** | **Geometric Scaler** | **Empirical Scaler** |
| IL1971 | Irvine and Lane 1971 (Table 4) | Le Houga Observatory, 12 inch and Boyden Observatory, 16 inch | 1963-1965 | Geometric Albedo | 0.359-1.064 | 0 | Disk integrated | 1 | 1.15 |
| C1979_c | Clark and McCord 1979 (Fig. 4) | Mauna Kea Observatory, 2.24 m | 2/2/1977 | Albedo | 0.664-2.503 | n/a | Center of Disk | 0.66 | 1.05 |
| C1979_l | Clark and Mccord 1979 (Fig. 5) | Mauna Kea Observatory, 2.24 m | 2/2/1977 | Albedo | 0.673-2.504 | n/a | West Limb | 1 | 0.87 |
| FL1979 | Fink and Larson 1979 (Fig. 2b) | Kitt Peak National Observatory, 4 m | 02/28/1971-03/02/1971 | Geometric albedo | 0.960-2.490 | 0 | Center of Disk | 1 | 1.2 |
| C1982 | Clarke et al. 1982 (Fig. 3) | International Ultraviolet Explorer | March-July 1980 | Geometric Albedo | 0.122-0.190 | n/a | Disk Averaged | 1 | 1 |
| K1998 | Karkoschka 1998 (Fig1) | European Southern Observatory, 1.52 m | 7/6/1995-7/10/1995 | Full-disk Albedo | 0.300-1.050 | 5.7 | Disk Averaged | 1 | 1 |
| R2017 | Roberge et al. 2017 (Fig. 2) | IRTF/ Telescope SpeX | n/a | Geometric Albedo | 1.002-2.497 | 0 | n/a | 1 | 1 |
| C2022 | Coulter et al. 2022 (Fig. 6) | Cassini/ VIMS | 08/02/2007 | I/F | 0.351-2.500 | 39.64 | Disk Integrated | 1.24 | 1.66 |

**Table 6:** Parameters describing each source used in the analysis of Saturn are reported. In order of appearance, the label used to describe the data (column 1), literature reference (2), mission or instruments used (3), date of the observations (4), units from the original publication (5),

wavelength range in μm (6), and phase angle in degrees at the time of the observations (7, also plotted in Figure 7A). Columns 8-10 report the geometry of the observation, the geometric scaling factor applied to normalize to a standard full-disk geometry (plotted in Figure 7B), and the empirical scaling factor used to match the geometric albedo data from the Karkoschka 1998 photometric reference spectrum (plotted in Figure 7C).

| Table 7: Titan | | | | | | | | | |
|---|---|---|---|---|---|---|---|---|---|
| **Source** | **Ref.** | **Mission/ Instrument** | **Date** | **Original units** | **Wavelengths (μm)** | **Phase Ang. (°)** | **Geometry** | **Geometric Scaler** | **Empirical Scaler** |
| Y1974 | Younkin 1974 (Table 3-2) | Mount Wilson Observatory, 60-inch reflector | 1/2/1974, 1/3/1974 | Geometric Albedo | 0.500-1.080 | 0 | Disk averaged | 1 | 0.8 |
| N1984 | Neff et al. 1984 (Table 4) | McDonald Observatory. 2.1 m | 5/17/1984-5/18/1984 | Geometric Albedo | 0.350-1.050 | 4.9 | Disk averaged | 1.004 | 1 |
| K1998 | Karkoschka 1998 (Fig1) | ESO, 1.52 m | 7/6/1995-7/10/1995 | Full-disk Albedo | 0.305-1.050 | 5.7 | Disk averaged | 1.005 | 1 |
| M1998 | McGrath et al 1998 Fig 7 | HST/ FOS | October 1991, August 1992 | Geometric Albedo | 0.180-0.330 | 0 | Disk-averaged | 1 | 1.2 |
| N2006_94, N2006_95 | Negrao et al. 2006 (Fig. 1, Fig. 5) | CFHT / FTS | 09/23/1994, 08/17/1995 | Geometric Albedo | 0.872-2.499, 1.040-2.495, | 0 | Disk averaged | 1 | 1 |
| N2006_m | Negrao et al. 2006 (Fig. 5) | | n/a | Geometric Albedo | 1.050-2.490 | 0 | Disk averaged | 1 | 0.9 |
| L2008 | Lavvas et al 2008 Fig. 19 | n/a (simulated) | n/a | Geometric Albedo | 0.153-5.930 | 0 | Disk averaged | 1 | 1.22 |

**Table 7:** Parameters describing each source used in the analysis of Titan are reported. In order of appearance, the label used to describe the data (column 1), literature reference (2), mission or instruments used (3), date of the observations (4), units from the original publication (5), wavelength range in μm (6), and phase angle in degrees at the time of the observations (7, also plotted in Figure 8A). Columns 8-10 report the geometry of the observation, the geometric scaling factor applied to normalize to a standard full-disk geometry (plotted in Figure 8B), and the empirical scaling factor used to match the geometric albedo data from the Karkoschka 1998 photometric reference spectrum (plotted in Figure 8C).

| Table 8: Uranus | | | | | | | | | |
|---|---|---|---|---|---|---|---|---|---|
| **Source** | **Ref.** | **Mission/ Instrument** | **Date** | **Original units** | **Wavelengths (μm)** | **Phase Ang. (°)** | **Geometry** | **Geometric Scaler** | **Empirical Scaler** |
| L1983 | Lockwood et al. 1983 (Table 4) | Lowell Observatory, 1.8 m | 06/18/1980, 04/24/1981, 06/17/1981, 06/18/1981 | Geometric Albedo | 0.332-0.763 | 0 | | 1 | 1.05 |
| N1984 | Neff et al. 1984 (Table 4) | McDonald Observatory. 2.1 m | 5/17/1981-5/18/1981 | Geometric Albedo | 0.350-1.051 | 0 | Disk integrated | 1 | 1.08 |
| W1986 | Wagener et al. 1986 (Table 4) | IUE Satellite | 10/2/1985 | Geometric Albedo | 0.205-0.334 | 0 | Disk integrated | 1 | 1 |
| K1998 | Karkoschka 1998 (Fig. 1) | European Southern Observatory, 1.52 m | 7/6/1995-7/10/1995 | Full-disk Albedo | 0.300-1.050 | 0.7 | Disk averaged | 1 | 1 |
| C1999 | Courtin 1999 (Fig. 12a) | HST/ FOS | 06/22/1992 | Geometric Albedo | 0.222-0.329 | 0.8 | Disk integrated | 1 | 1 |
| S2015 | Schmude et al. 2015 (Table 9) | multiple | n/a | Geometric Albedo | 0.360-0.798 | 0 | Disk integrated | 1 | 1 |
| R2017 | Roberge et al. 2017 (Fig. 2) | IRTF/SpeX | n/a | Geometric Albedo | 1.002-2.497 | 0 | Disk integrated | 1 | 1 |
| I2022_HST | Irwin et al. 2022 (Figure 1) | HST/ STIS | 08/19/2002 | I/F | 0.302-0.999 | 0.04 | Central disk to limb | 1 | 0.95 |
| I2022_IRTF | Irwin et al. 2022 (Figure 1) | IRTF/SpeX | 05/18/2000 | I/F | 0.807-2.500 | n/a | n/a | 1 | 1 |

**Table 8:** Parameters describing each source used in the analysis of Uranus are reported. In order of appearance, the label used to describe the data (column 1), literature reference (2), mission or instruments used (3), date of the observations (4), units from the original publication (5),

wavelength range in μm (6), and phase angle in degrees at the time of the observations (7). Columns 8-10 report the geometry of the observation, the geometric scaling factor needed to normalize to a standard full-disk geometry, and the empirical scaling factor used to match the geometric albedo data from the Karkoschka 1998 photometric reference spectrum (plotted in Figure 9B).

| Table 9: Neptune | | | | | | | | | |
|---|---|---|---|---|---|---|---|---|---|
| **Source** | **Ref.** | **Mission/ Instrument** | **Date** | **Original units** | **Wavelengths (μm)** | **Phase Ang. (°)** | **Geometry** | **Geometric Scaler** | **Empirical Scaler** |
| FL1979 | Fink and Larson 1979 (Fig. 2a) | Kitt Peak National Observatory, 4 m | 02/28/1975-03/02/1975 | Geometric Albedo | 0.939-2.450 | 0 | | 1 | 0.70 |
| N1984 | Neff et al. 1984 | McDonald Observatory. 2.1 m | 5/17/1981-5/18/1981 | Geometric Albedo | 0.350-1.051 | 0 | Disk integrated | 1 | 1.25 |
| W1986 | Wagener et al. 1986 (Table 4) | IUE Satellite | 10/2/1985-10/4/1985 | Geometric Albedo | 0.205-0.334 | 0 | Disk integrated | 1 | 1.1 |
| K1998 | Karkoschka 1998 (Fig. 1) | European Southern Observatory, 1.52 m | 7/6/1995-7/10/1995 | Full-disk Albedo | 0.300-1.050 | 0.3 | Disk averaged | 1 | 1.1 |
| C1999 | Courtin 1999 (Fig. 2) | HST/ FOS | 06/22/1992, 08/19/1992 | Geometric Albedo | 0.221-0.330 | 1.2 | | 1 | 1.1 |
| R2017 | Roberge et al. 2017 (Fig. 2) | IRTF/SpeX | n/a | Geometric Albedo | 1.040-2.497 | 0 | Disk integrated | 1 | 1 |
| I2022_HST | Irwin et al. 2022 (Figure 1) | HST/ STIS | 08/03/2003 | I/F | 0.301-0.999 | 0.04 | Central disk to limb | 1 | 1 |
| I2022_IRTF | Irwin et al. 2022 (Figure 1) | IRTF/SpeX | 06/30/2000 | I/F | 1.004-2.496 | n/a | n/a | 1 | 1 |

**Table 9:** Parameters describing each source used in the analysis of Neptune are reported. In order of appearance, the label used to describe the data (column 1), literature reference (2), mission or instruments used (3), date of the observations (4), units from the original publication (5), wavelength range in μm (6), and phase angle in degrees at the time of the observations (7). Columns 8-10 report the geometry of the observation, the geometric scaling factor needed to normalize to a standard full-disk geometry, and the empirical scaling factor used to match the geometric albedo data from Irwin et al. 2022 (plotted in Figure 10B).

**Table 10: Planets colorimetry, absolute magnitudes and B-V color index**

| Planet | Rh (au) [a] | D (km) [b] | RGB [c] | Hvis (mag) [d] | B-V Color Index | | Ageo | | |
|---|---|---|---|---|---|---|---|---|---|
| | | | | This work | This work | Mallama et al. 2017 | This work Geye | This work V Johnson | Mallama et al. 2017 |
| Mercury | 0.39 | 4879 | 1<br>0.98<br>0.87 | -0.71 | 0.86 | 0.97 | 0.133 | 0.143 | 0.142 |
| Venus | 0.72 | 12104 | 098<br>1<br>0.83 | -4.37 | 0.92 | 0.92 | 0.683 | 0.678 | 0.689 |
| Earth | 1.0 | 12756 | 0.77<br>0.82<br>1 | -3.23 | 0.40 | 0.47 | 0.218 | 0.213 | 0.434 |
| Mars | 1.52 | 6779 | 1<br>0.83<br>0.31 | -1.65 | 1.82 | 1.56 | 0.132 | 0.177 | 0.17 |
| Jupiter | 5.2 | 142796 | 0.99<br>1<br>0.87 | -9.45 | 0.83 | 0.88 | 0.528 | 0.524 | 0.53 |
| Saturn | 9.54 | 120660 | 1<br>0.98<br>0.71 | -9.05 | 1.06 | 1.07 | 0.477 | 0.505 | 0.499 |
| Titan | 9.54 | 5149.5 | 1<br>0.94<br>0.58 | -1.36 | 1.24 | | 0.208 | 0.232 | |
| Uranus | 19.0 | 51118 | 0.82<br>0.90<br>1 | -7.06 | 0.53 | 0.50 | 0.544 | 0.452 | 0.488 |
| Neptune | 30.1 | 49523 | 0.71<br>0.70<br>1 | -6.81 | 0.38 | 0.39 | 0.478 | 0.381 | 0.442 |

**Table 10:** For each planet, we provide the heliocentric distance (AU) and diameter (km), followed by the results of our analysis: RGB color values, visual magnitude ($H_{Vis}$), and the (B–V) color index from both this study and Mallama et al. (2017). Geometric albedo values are presented as derived from this study using the green eye filter and Johnson V band, alongside the corresponding Johnson V band albedo values from Mallama et al. (2017).

| Table 11: Exoplanet analogs | | | | | |
|---|---|---|---|---|---|
| Object | Semi Major Axis (au) | Analog Planet | Radius Ratio | Blackbody Temperature (K) | Mass (kg) |
| Proxima Centauri (star) | n/a | n/a | 0.15 $R_{Sun}$ | 2900 | 2.43e29 |
| Proxima Centauri b | 0.04856 | Earth | 1.02 $R_{Earth}$ | 229.4 | 6.39e24 |
| Proxima Centauri c | 1.48 | Mars | 2.01 $R_{Earth}$ | 44.2 | 3.46e25 |
| Proxima Centauri d | 0.02885 | Mercury | 0.81 $R_{Earth}$ | 316.7 | 1.55e24 |
| HD 219134 (star) | n/a | n/a | 0.78 $R_{Sun}$ | 4817 | 1.61e30 |
| HD 219134 b | 0.0388 | Mercury | 1.54 $R_{Earth}$ | 1012 | 2.74e25 |
| HD 219134 c | 0.0653 | Mercury | 1.46 $R_{Earth}$ | 781.8 | 2.53e25 |
| HD 219134 d | 0.237 | Venus | 1.61 $R_{Earth}$ | 311.8 | 9.66e25 |
| HD 219134 f | 0.1463 | Venus | 1.31 $R_{Earth}$ | 396.7 | 4.36e25 |
| HD 219134 g | 0.375 | Neptune | 3.31 $R_{Earth}$ | 298.4 | 6.45e25 |
| HD 219134 h | 3.11 | Jupiter | 6.81 $R_{Earth}$ | 114.3 | 4.56e26 |

**Table 11:** Parameters used in the analysis of exoplanets in the Proxima Centauri and HD 219134 systems. For each planet, the following quantities are listed: semi-major axis (AU), corresponding solar system analog from this study, planet-to-Earth radius ratio, estimated blackbody temperature (K), and planetary mass (kg).

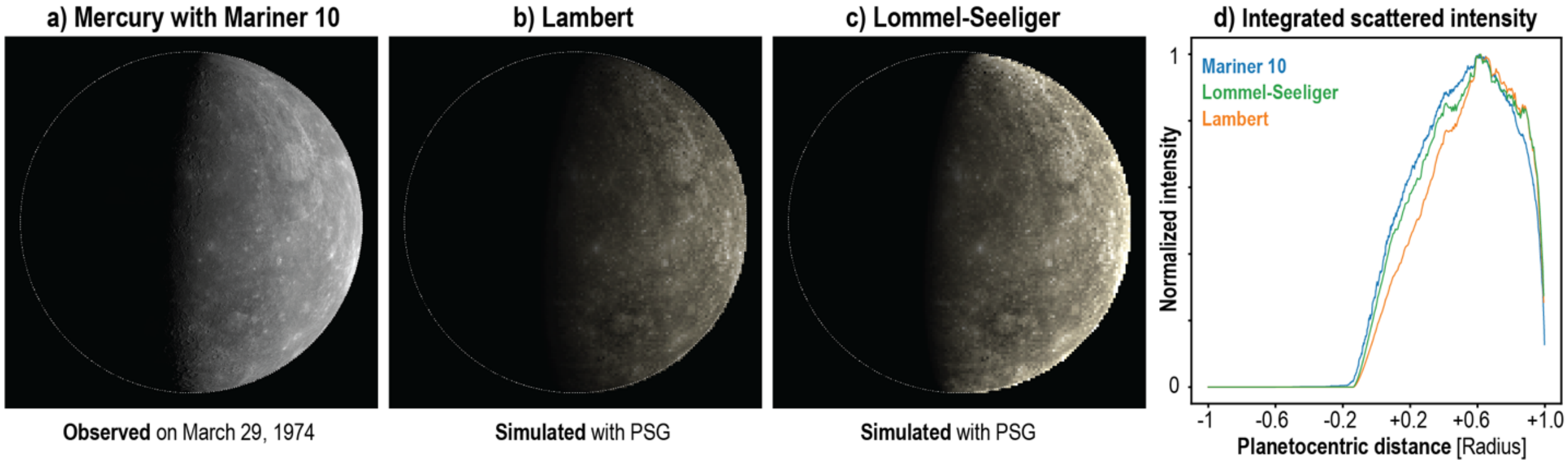

**Figure 1:** Comparison of single-scattering surface models to a Mariner 10 image. a) Mercury is shown in quadrature as observed by the Mariner 10 spacecraft on March 29th, 1974. b) Reflected sunlight is simulated on the planet for the observations using a Lambert scattering model. c) Mercury surface fluxes synthesized using a Lommel-Seeliger scattering model. d) Integrated scattering intensities are shown for the Mariner 10 observation, Lommel-Seeliger model, and the Lambert model along the planet's longitudinal axis.

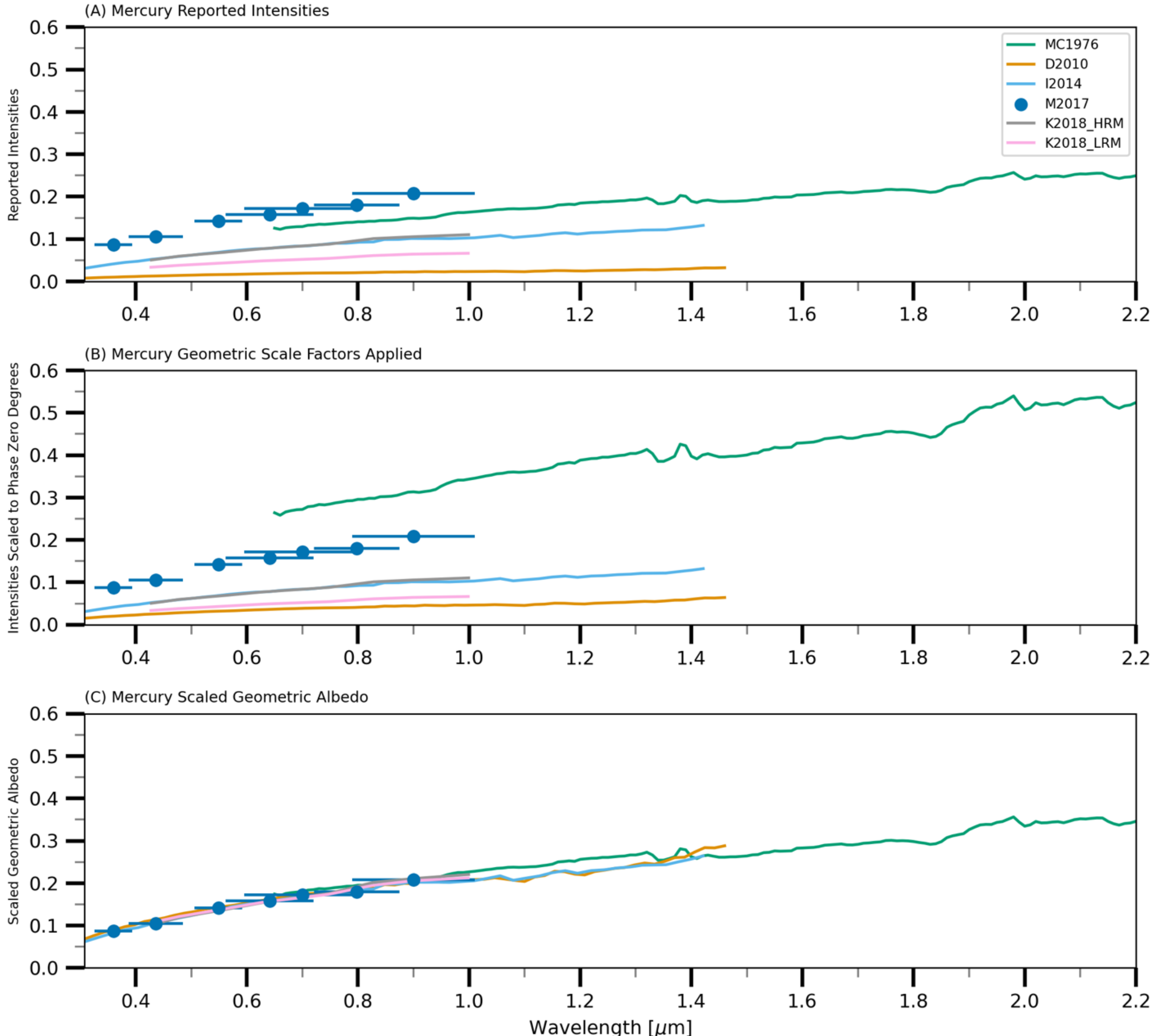

**Figure 2:** Reflectance spectra of Mercury are plotted and scaled, from top to bottom, to provide consistent calibrated data. In each of the panels, the solid colored lines represent spectroscopic observations, and the colored dots correspond to photometric filter measurements. Observations from Mallama et al. 2017 (M2017) were selected as the reference to scale the other sources to. *Panel A*: The intensities of the reflectance values found in the original publications (MC1979: McCord and Clark 1979 , D2010: Domingue et al. 2010, I2014: Izenberg et al. 2014, M2017: Mallama et al. 2017, K2018_HRM: Klima et al. 2018 (High Reflectance Material), K2018_LRM: Klima et al. 2018 (Low Reflectance Material). *Panel B:* Data are corrected to a phase angle of zero using the Lommel Seeliger phase coefficients. Geometric scale factors are applied to MC1979 and D2010. *Panel C*: Data in panel B are scaled to match the geometric albedo values from M2017. Correction factors are applied to Y1974, M1998, and L2008.

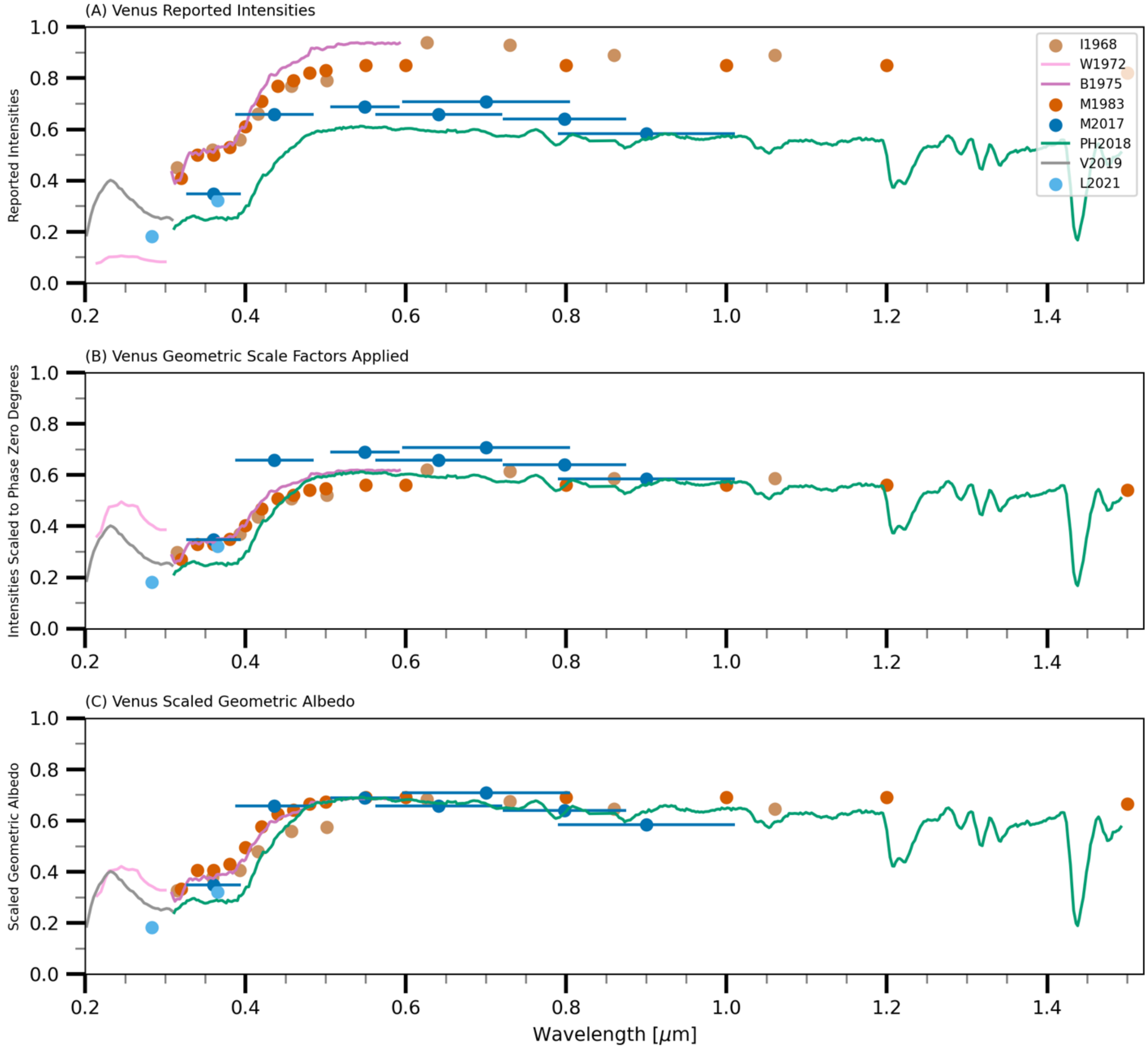

**Figure 3:** Reflectance spectra of Venus are plotted and scaled, from top to bottom, to provide consistent calibrated data. In all the panels, the full colored lines represent spectroscopic observations, the colored dots correspond to photometric filter measurements, and the dashed line refers to synthetic reflectance spectra. Observations from Mallama et al. 2017 (M2017) were selected as the reference to scale the other sources to. *Panel A*: The intensities of the reflectance values found in the original publications (I1968: Irvine 1968, W1972: Wallace et al. 1972, B1975: Barker et al. 1975, M1983: Moroz 1983, R2017: Roberge et al. 2017, PH2018: Perez-Hoyos et al. 2018, V2019: Vlasov et al. 2019, L2021: Lee et al. 2021). *Panel B*: Data are scaled to a phase angle of zero using the Lambertian phase coefficients. Geometric scale factors are applied to I1968, W1972, B1975, and M1983. *Panel C*: Data in panel B are scaled to match the visual albedo, 0.689, reported in Mallama et al. 2017 at 0.549 μm. Correction factors are applied to I1968, W1972, B1975, M1983, and PH2018.

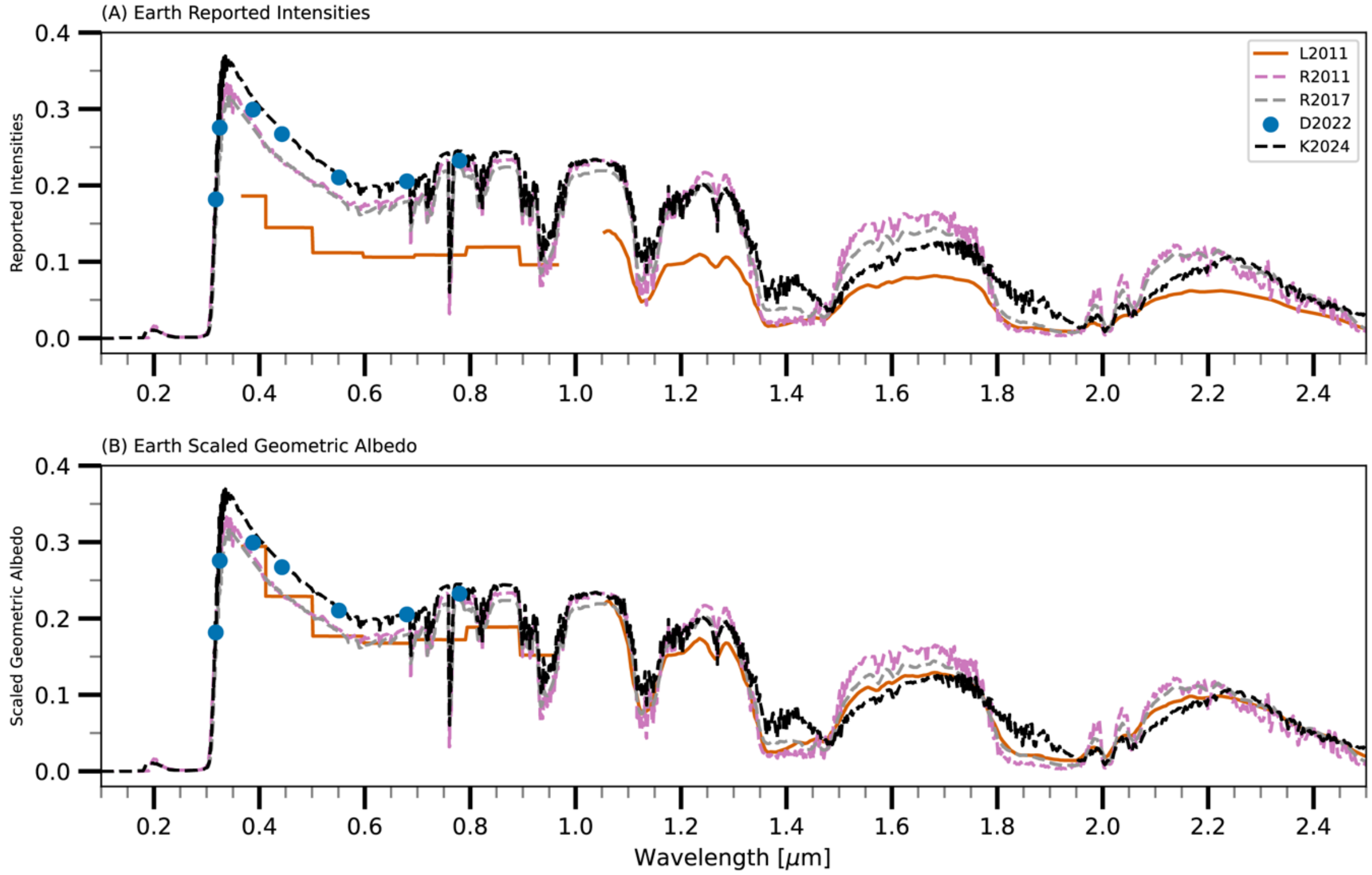

**Figure 4:** Reflectance spectra of Earth are plotted and scaled, from top to bottom, to provide consistent calibrated data. In each of the panels, the full colored lines represent spectroscopic observations, the colored dots correspond to photometric measurements, and the dashed lines refer to synthetic reflectance spectra. *Panel A*: The intensities of the reflectance values found in the original publications (L2011: Livengood et al. 2001, R2011: Robinson et al. 2011, R2017: Roberge et al. 2017, D2022: DSCOVR 2022, K2024: Kofman et al. 2024). *Panel B:* Data are corrected to a phase angle of zero using the Lambertian phase coefficients. A geometric scale factor is applied to L2011. The K2024 model was specifically calculated for the exact conditions and geometry of D2022.

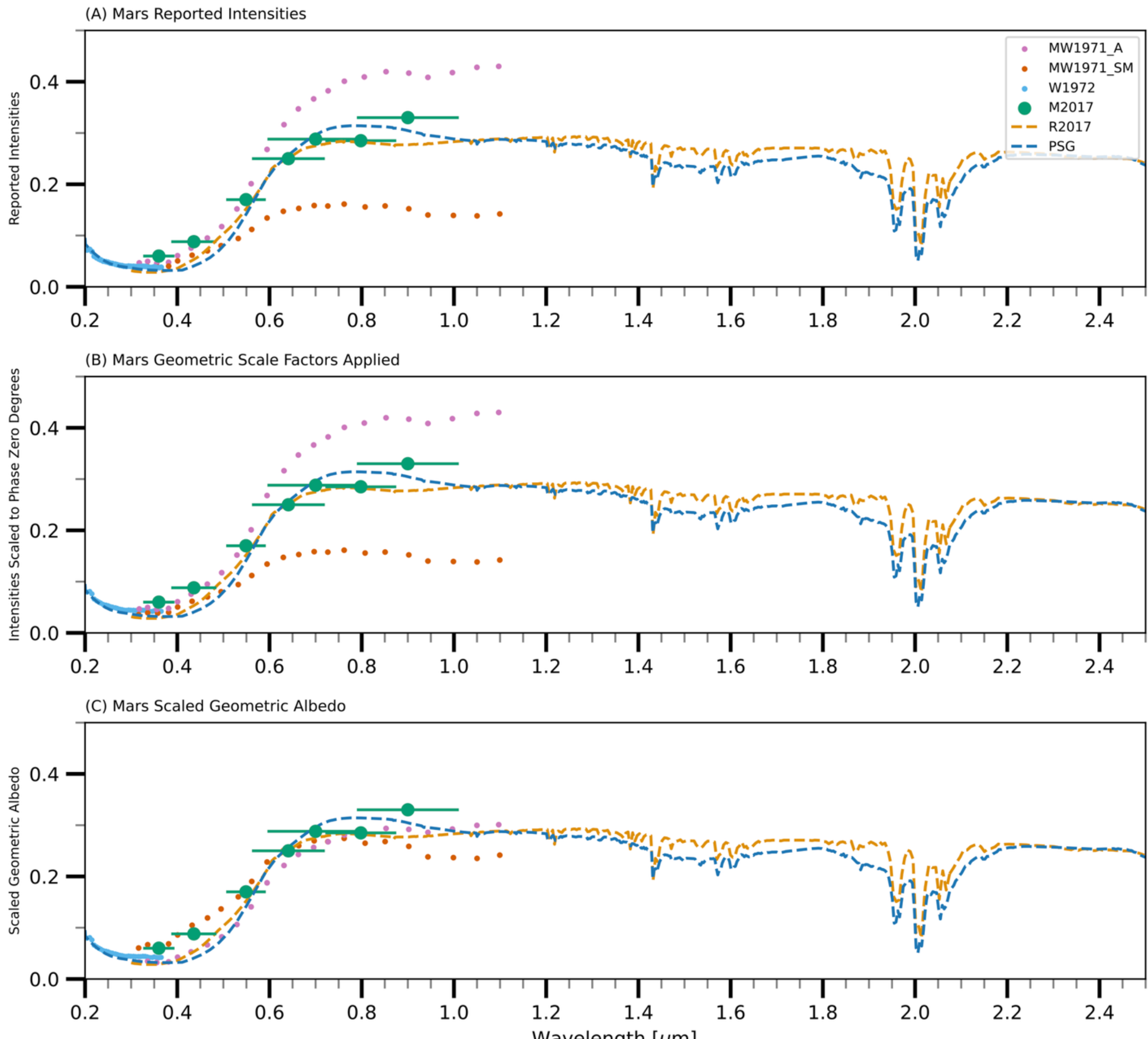

**Figure 5:** Reflectance spectra of Mars are plotted and scaled, from top to bottom, to provide consistent calibrated data. In all the panels, the full colored lines represent spectroscopic observations, the colored dots correspond to photometric filter measurements, and the dashed line refers to synthetic reflectance spectra. Data from Mallama et al. 2017 (M2017) is our chosen reference spectrum. *Panel A*: The intensities of the reflectance values found in the original publications MW1971_SM: Mccord and Westphal 1971 (Syrtis Major), MW1971_A: Mccord and Westphal 1971 (Arabia), W1972: Wallace et al. 1972, R2017: Roberge et al. 2017, PSG: Villanueva et al. 2022). *Panel B:* Data are corrected to a phase angle of zero using the Lambertian phase coefficients. *Panel C*: Data in panel B are scaled to match the continuum reported in M2017 (correction factors reported in Table 4).

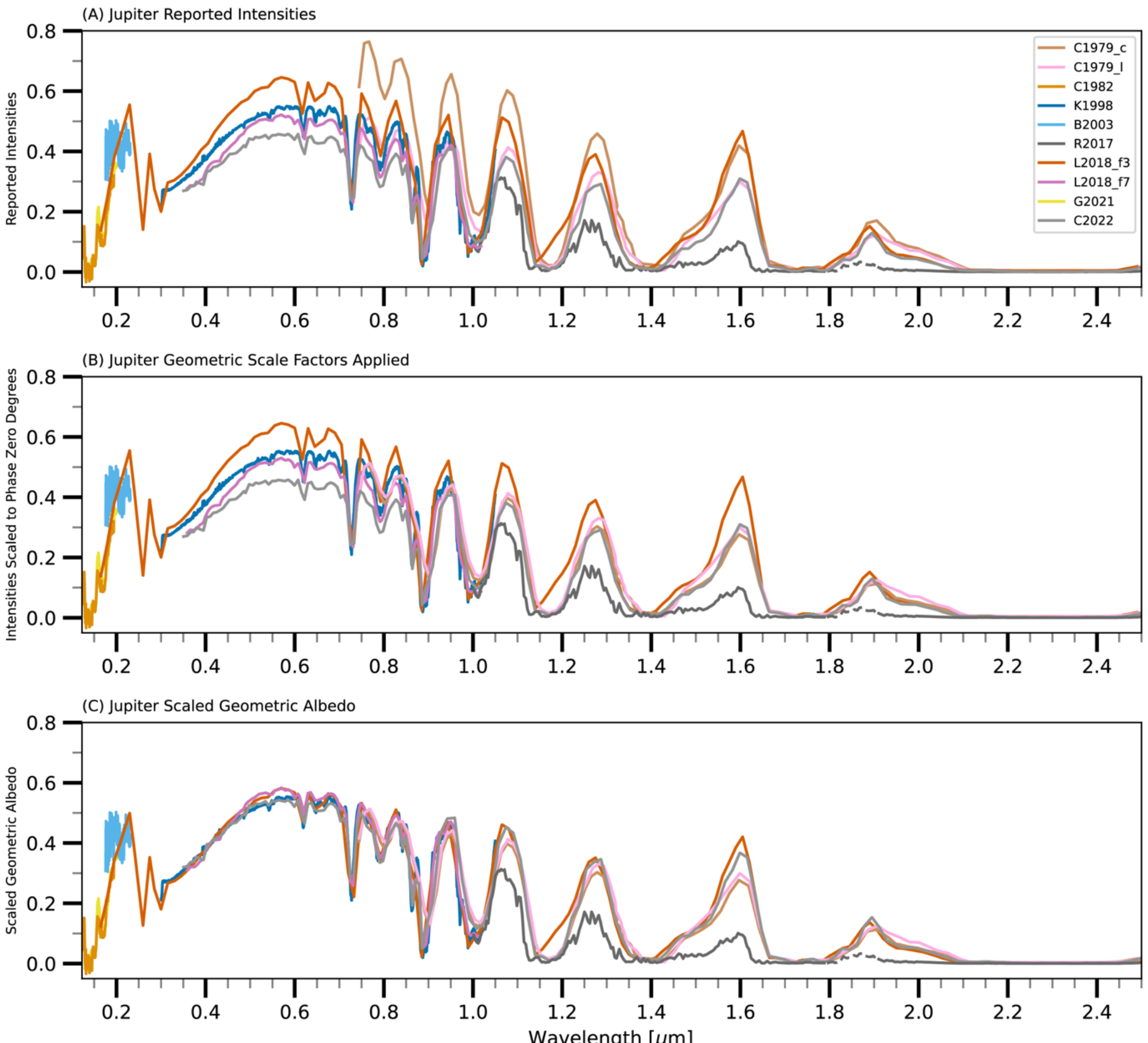

**Figure 6:** Reflectance spectra of Jupiter are plotted and scaled, from top to bottom, to provide consistent calibrated data. In all the panels the full colored lines represent spectroscopic observations and dashed lines display synthetic data. Data from Karkoschka 1998 is the chosen reference spectrum used to scale the other sources to. *Panel A*: The intensities of the reflectance values found in the original publications (C1979_c: Clark and McCord 1979 central disk data, C1979_l: Clark and McCord limb data, C1982: Clarke et al. 1982, K1998: Karkoschka 1998, B2003: Betremieux et al. 2003, R2017: Roberge et al. 2017, L2018_f3: Li et al. 2018 Figure 3, L2018_f7: Li et al. 2018 Supplementary Figure 7, G2021: Giles et al. 2021, C2022: Coulter et al. 2022). *Panel (B):* Data are corrected to a phase angle of zero using the Lambertian phase coefficients. Geometric scale factors are applied to C1979_c, K1998, L2018_f7. *Panel (C)*: Data in panel B are scaled to match the continuum reported in Karkoschka 1998. Correction factors are applied to L2018_f3, L2018_f7, and C2022.

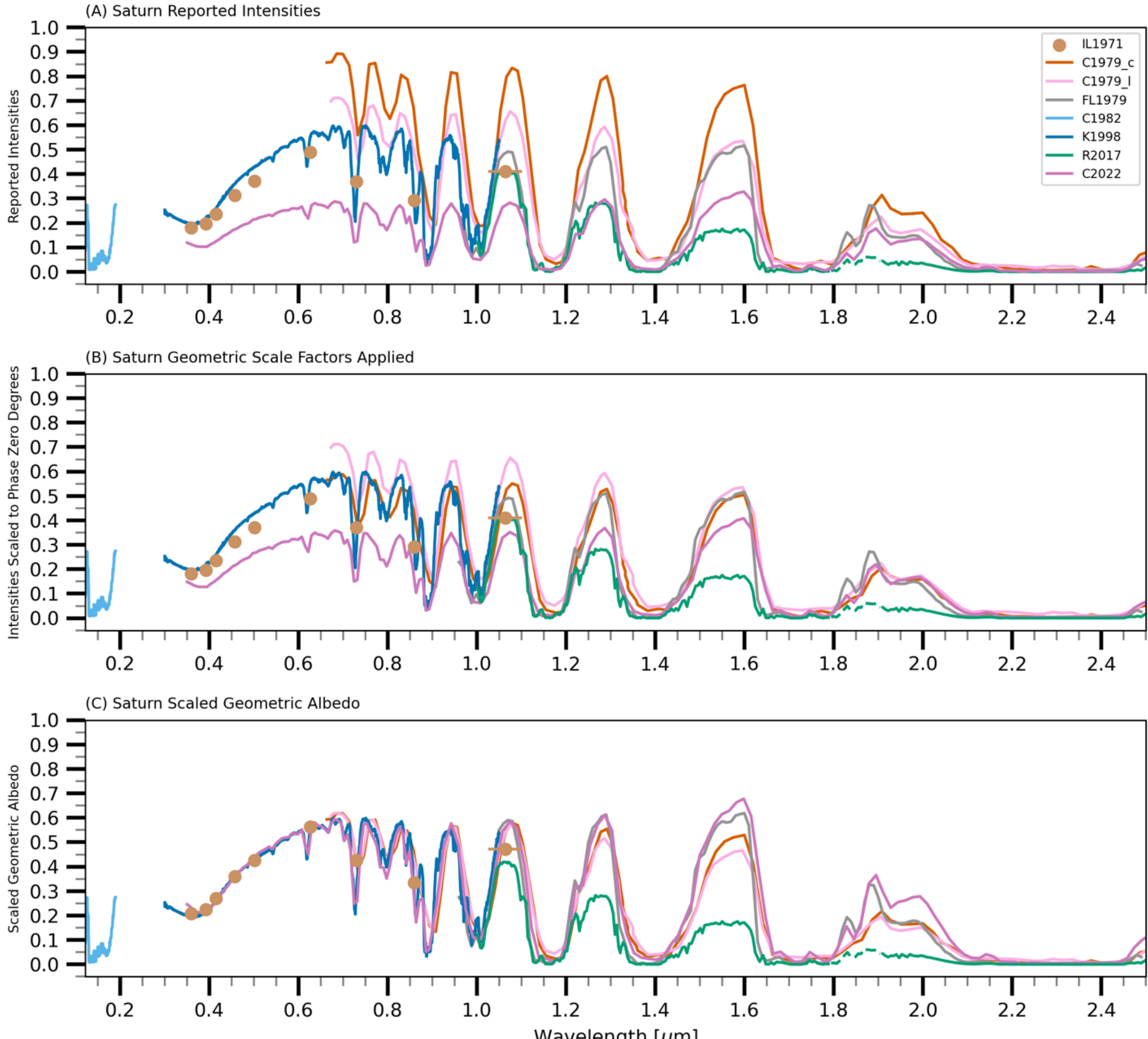

**Figure 7:** Reflectance spectra of Saturn are plotted and scaled, from top to bottom, to provide consistent calibrated data. In all the panels, the full colored lines represent spectroscopic observations, the colored dots correspond to photometric filter measurements, with their filter widths shown as horizontal lines. Data from Karkoschka 1998 (K1998) is our chosen reference spectrum used to scale the other sources to. *Panel A:* The intensities of the reflectance values found in the original publications (IL1971: Irvine and Lane 1971, C1979_c: Clark and McCord 1979 (center of disk), C1979_l, Clark and McCord 1979 (limb), FL1979: Fink and Larson 1979, C1982: Clarke et al. 1982, K1998: Karkoschka 1998, R2017: Roberge et al. 2017, C2022: Coulter et al. 2022. *Panel B:* Data are corrected to a phase angle of zero using the Lambertian phase coefficients. Geometric scale factors are applied to C1979_c and C2022. *Panel C:* Data in panel B are scaled to match the continuum reported in Karkoschka 1998. Correction factors are applied to IL1971, C1979_c, C1979_l, FL1979, and C2022.

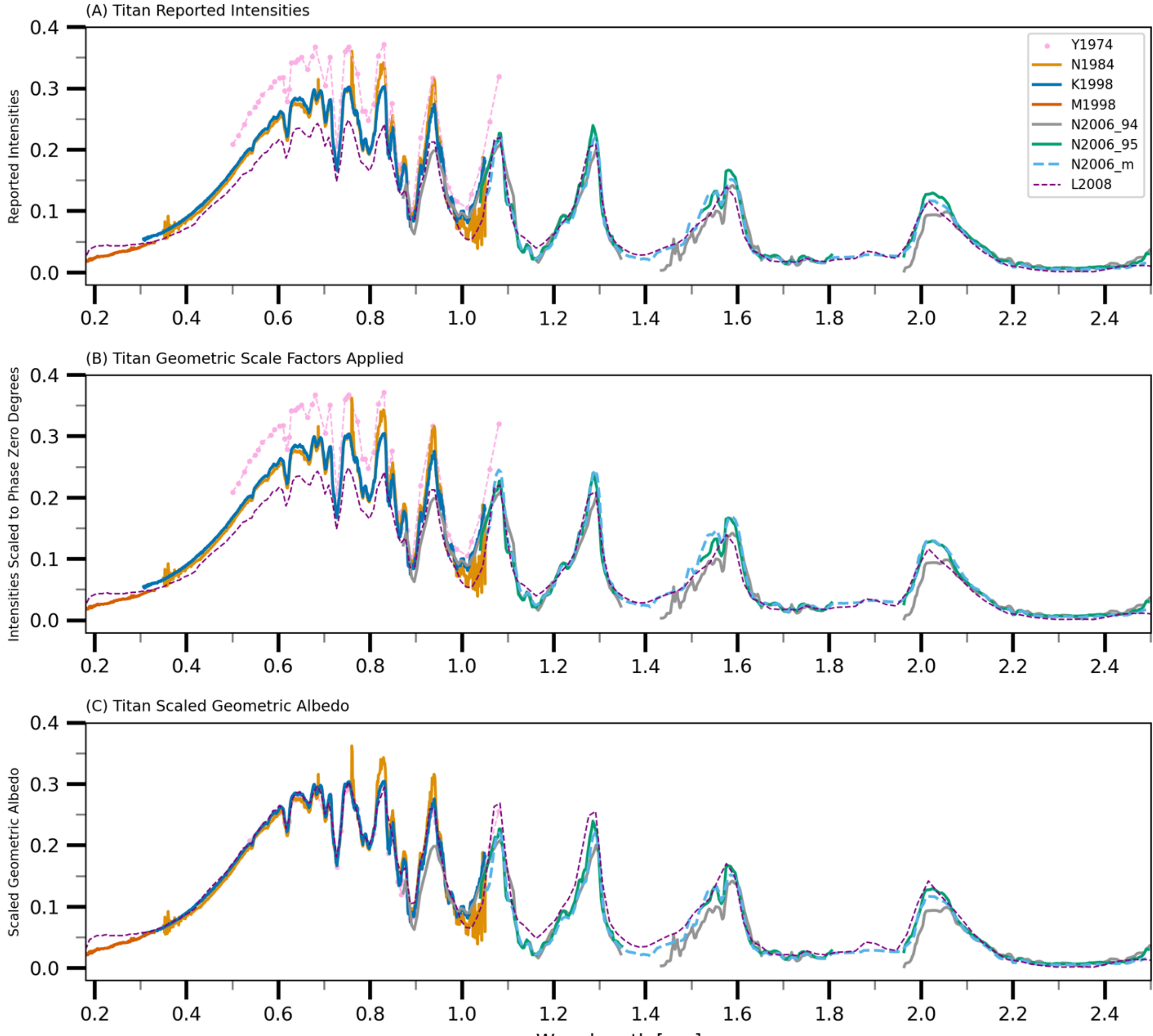

**Figure 8:** Reflectance spectra of Titan are plotted and scaled, from top to bottom, to provide consistent calibrated data. In all the panels, the full colored lines represent spectroscopic observations, the colored dots correspond to photometric filter measurements, and the dashed line refers to synthetic reflectance spectra. Data from Karkoschka 1998 (K1998) is our chosen reference spectrum used to scale the other sources to. *Panel A*: The intensities of the reflectance values found in the original publications (Y1974: Younkin 1974, N1984: Neff et al. 1984, K1998: Karkoschka 1998, M1998: McGrath et al. 1998, N2006_94: Negrao et al. 2006 (1994 observations), N2006_95: Negrao et al. 2006 (1995 observations), N2006_m: Negrao et al. 2006 (model data), L2008: Lavvas et al. 2008). *Panel B:* Data are corrected to a phase angle of zero using the Lambertian phase coefficients. Geometric scale factors are applied to N1984 and K1998. *Panel C*: Data in panel B are scaled to match the continuum reported in Karkoschka 1998. Correction factors are applied to Y1974, M1998, and L2008.

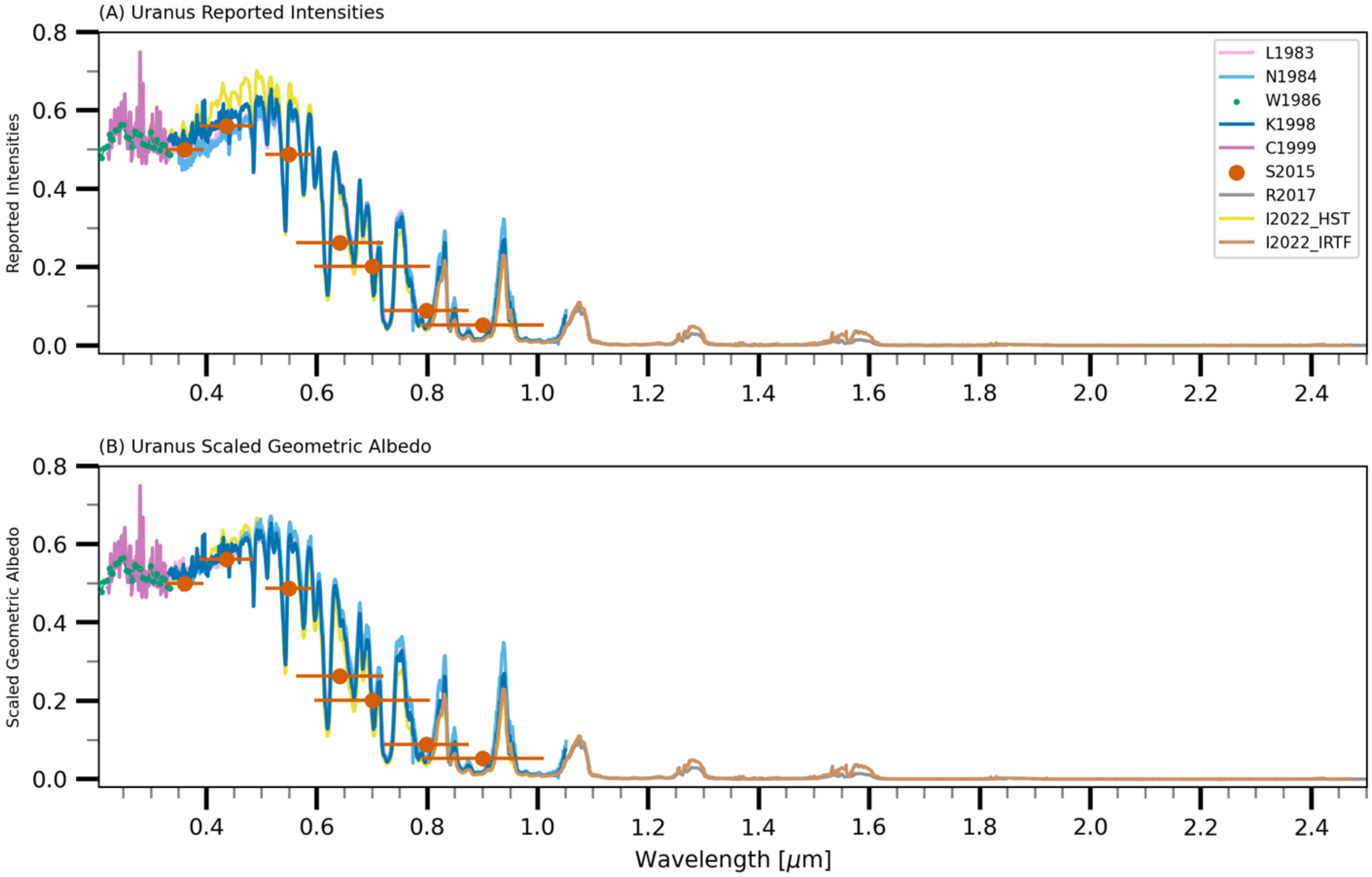

**Figure 9:** Reflectance spectra of Uranus are plotted and scaled, from top to bottom, to provide consistent calibrated data. In all the panels, the full colored lines represent spectroscopic observations, the colored dots correspond to photometric filter measurements, with their filter widths shown as horizontal lines. Data from Karkoschka 1998 (K1998) is our chosen reference spectrum used to scale the other sources to. *Panel A:* The intensities of the reflectance values found in each of the original publications (L1973: Lockwood et al. 1973, N1984: Neff et al. 1984, W1986: Wagener et al. 1986, K1998: Karkoschka 1998, C1999: Courtin 1999, S2015: Schmude et al. 2015, R2017: Roberge et al. 2017, I2022_HST: Irwin et al. 2022 (HST data), I2022_IRTF: Irwin et al. 2022 (IRTF data) are displayed. *Panel B:* Data in panel A are scaled to match the continuum reported in Karkoschka 1998. Correction factors are applied to L1983, N1984, and I2022_HST.

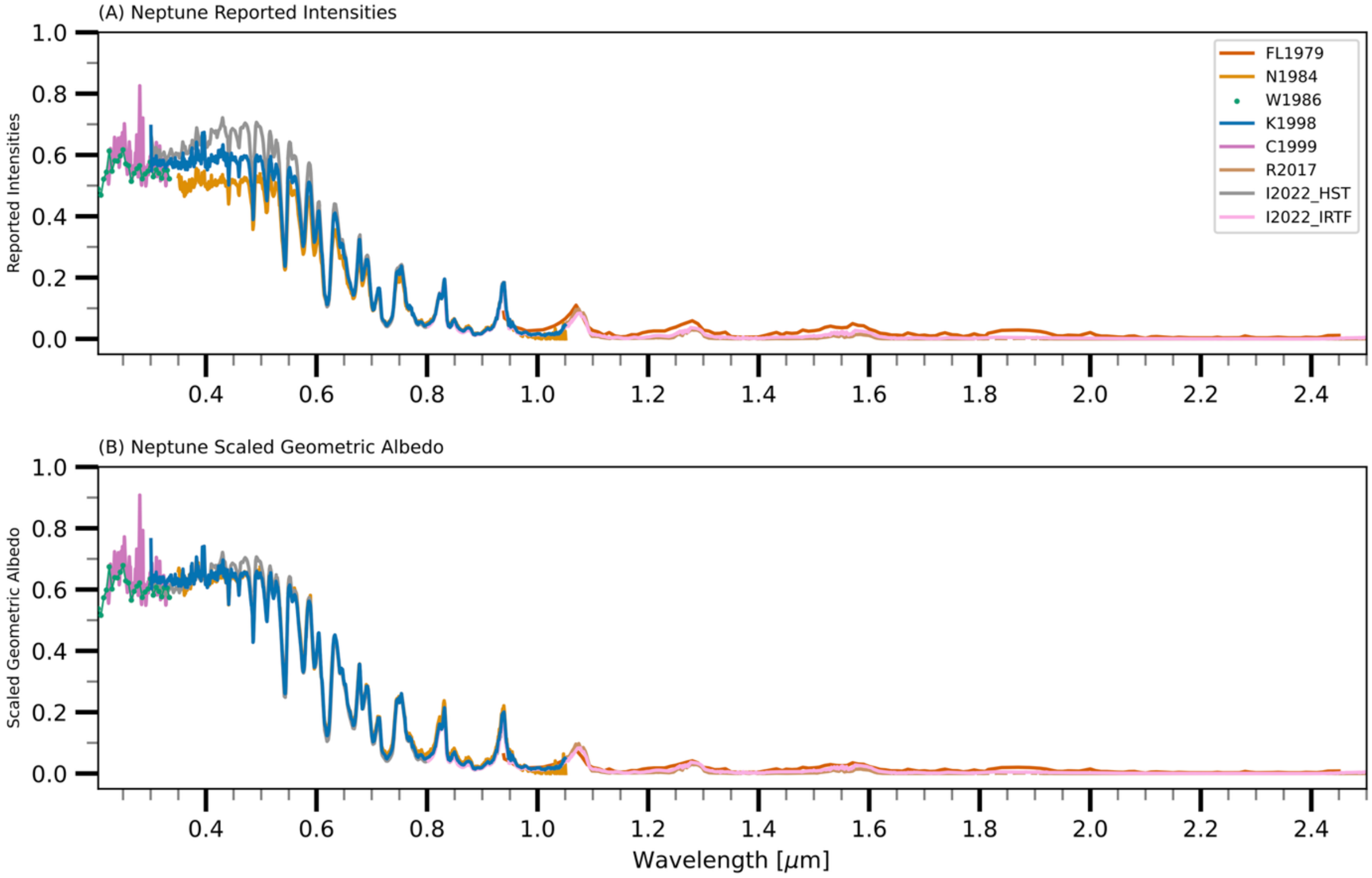

**Figure 10:** Reflectance spectra of Neptune are plotted and scaled, from top to bottom, to provide consistent calibrated data. In all the panels, the full colored lines represent spectroscopic observations, the colored dots correspond to photometric filter measurements, with their filter widths shown as horizontal lines. Most recent data from Irwin et al. 2022 are defined as the reference spectrum used to scale the other sources to. *Panel A:* The intensities of the reflectance values found in each of the original publications (FL1979: Fink and Larson 1979, N1984: Neff et al. 1984, W1986: Wagener et al. 1986, K1998: Karkoschka 1998, C1999: Courtin 1999, R2017: Roberge et al. 2017, I2022_HST: Irwin et al. 2022 (HST observations), I2022_IRTF: Irwin et al. 2022 (IRTF observations) are displayed. *Panel B:* Data in panel A are scaled to match the continuum reported in Irwin et al. 2022. Correction factors are applied to FL1979, N1984, W1986, K1998 and C1999.

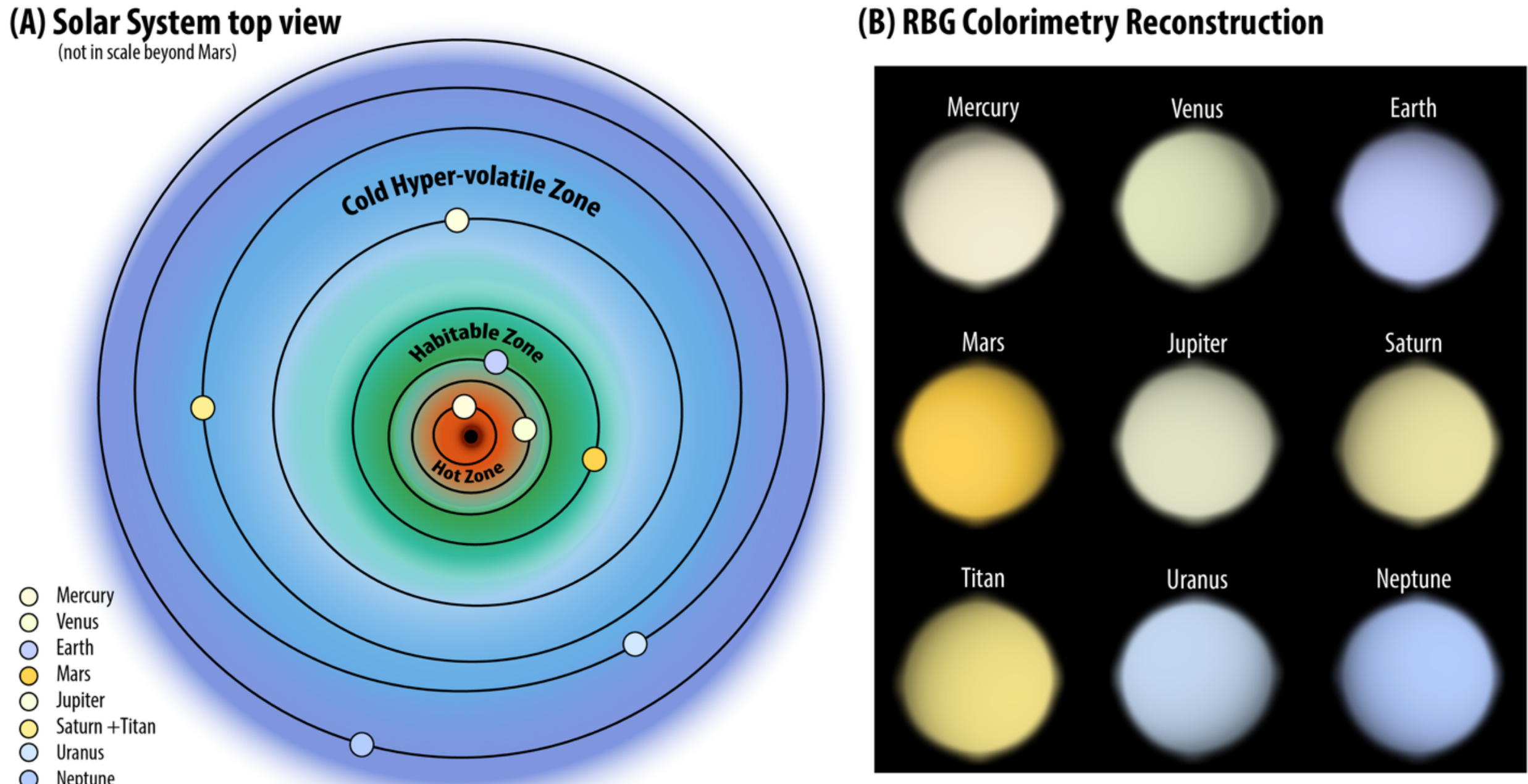

**Figure 11:** Panel A: Top-view of the Solar System and true color of the planets. The orbital distances are in scale until Mars, while the outer planets are not in scale. The Hot Zone, Habitable Zone, and Cold Hyper-volatile Zone are highlighted, and planets true color are shown. Panel B: Results of the RGB color reconstruction for each planet. The planetary colors are displayed for unresolved, disk-integrated spectra convolved with the digitized filter as reported in Irwin et al. 2024, to reproduce the colors as viewed with the human eye.

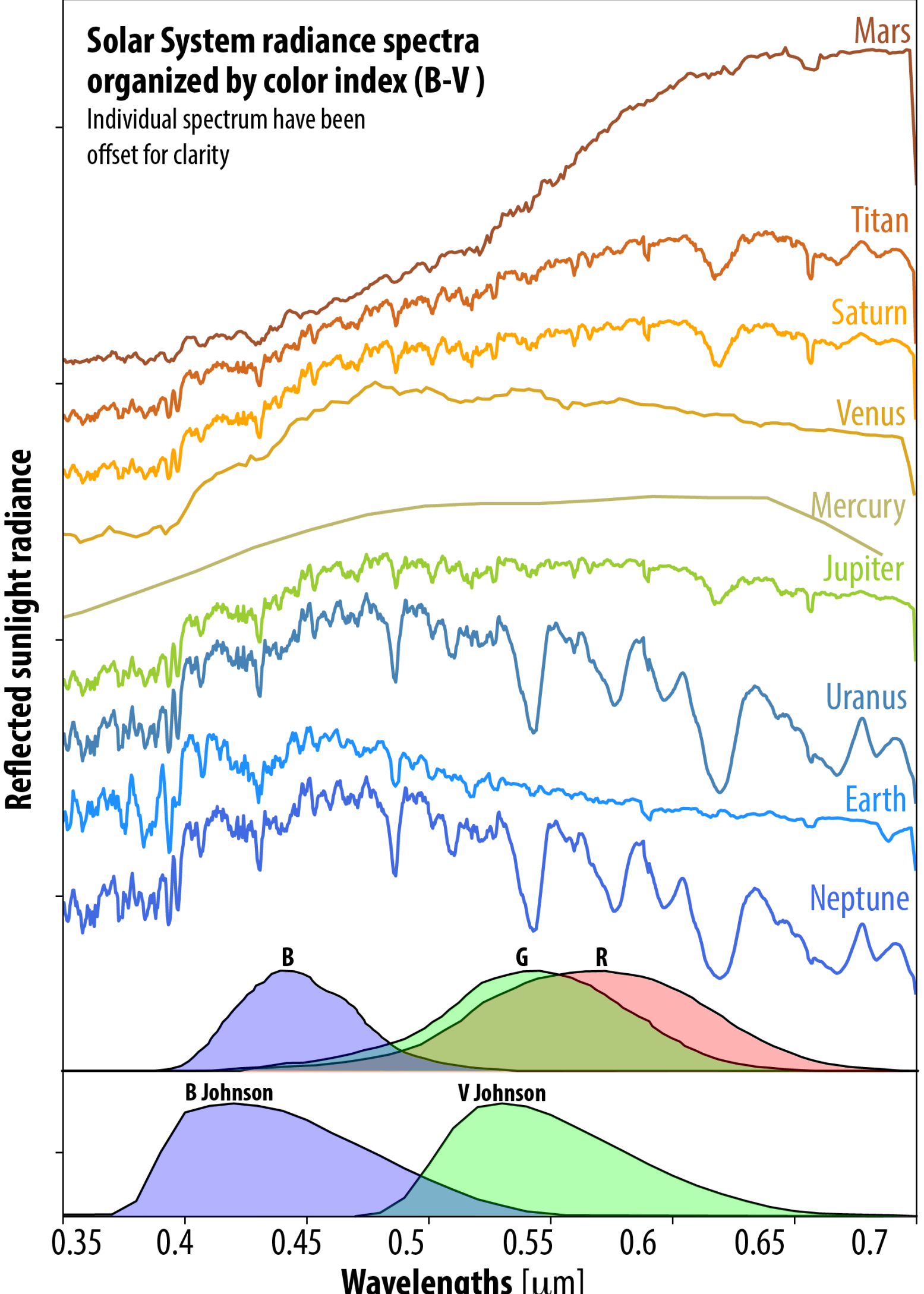

**Figure 12**: Spectra of the solar system planets and Titan are shown from top to bottom, ordered by decreasing (B-V) color index. The data have been smoothed and arranged to illustrate how the spectral slope in the visible range influences perceived color. The blue (B), green (G), red (R), and Johnson B and V filters are adopted in this analysis and overlaid at the bottom of the figure.

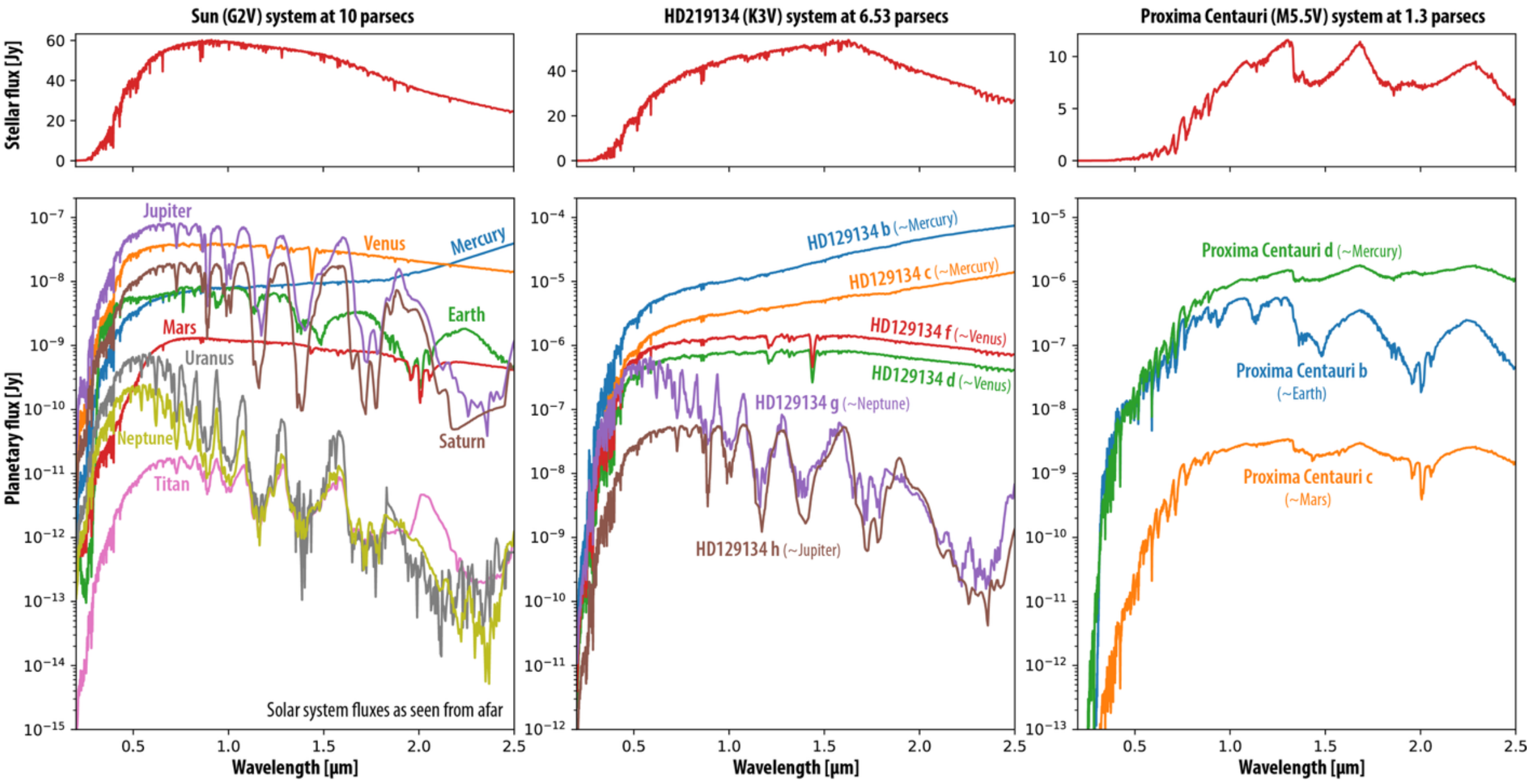

**Figure 13:** Left: Spectral flux densities are displayed for each of the solar system planets and Titan using the reference spectra created in this study for the solar system as viewed from 10 pc away. The planetary spectra are plotted at quadrature (phase angle of 90). The stellar flux for the Sun (spectral type: G2V) is displayed across the same wavelength range as each of the planets in our study. Middle: Spectral flux densities are displayed for each of the planets in the HD 219134 system by adopting the reference spectrum of each solar system analog at a phase angle of 0. The results are shown at the true distance of the planetary system, 6.54 parsecs. The stellar flux for HD 219134 (spectral type: K3V) is displayed across the same wavelength range as each of the planets in the study. Right: Spectral flux densities are displayed for each of the planets in the Proxima Centauri system by adopting the reference spectrum of each solar system analog at a phase angle of 0. The results are shown at the true distance of the planetary system, 1.30 parsecs. The stellar flux for Proxima Centauri (spectral type: M5.5V) is displayed across the same wavelength range as each of the planets in the study.

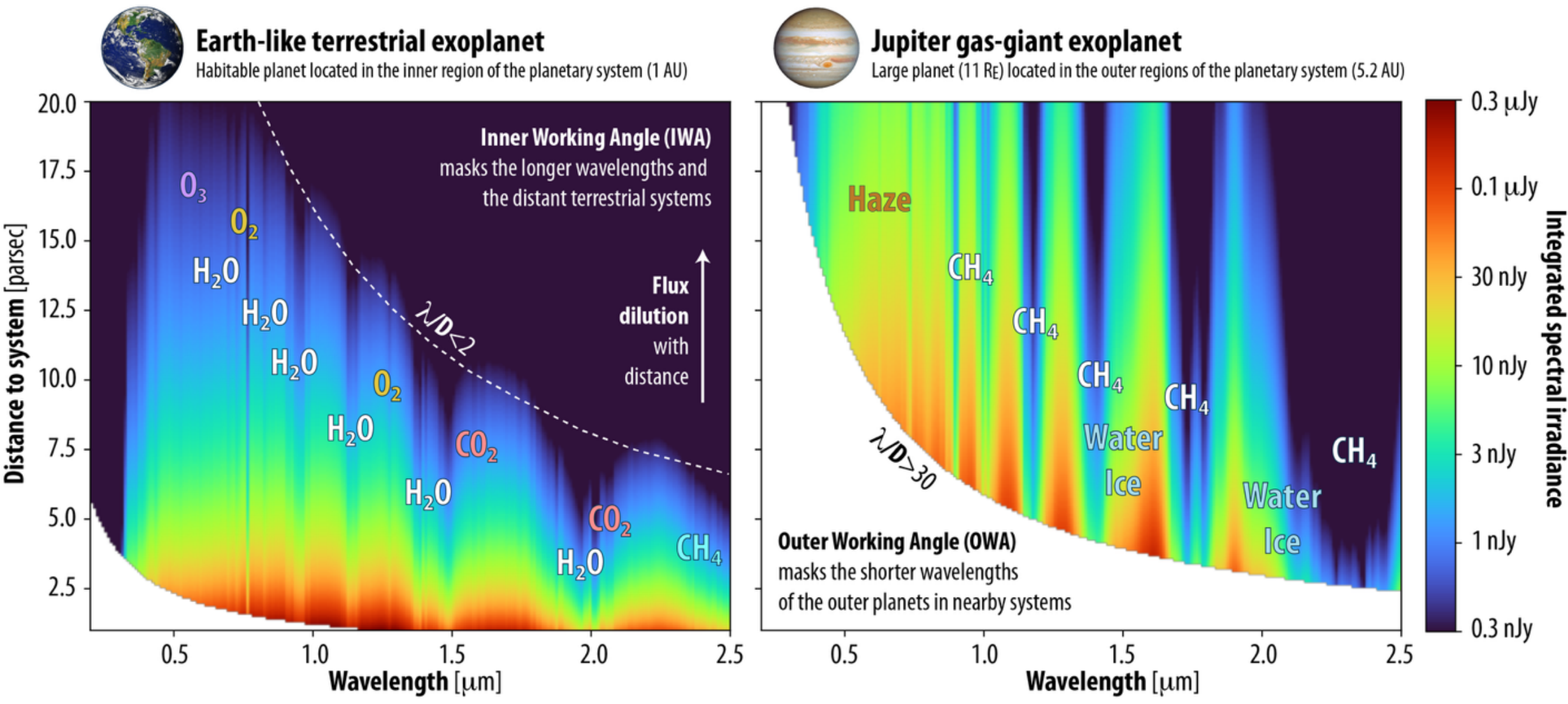

**Figure 14:** [Left] Simulated atmospheric absorption features for an Earth-like terrestrial exoplanet orbiting at 1 au from its host star. The spectrum spans the wavelength range from 0.2 to 2.5 μm (as focused on its this work), and the system distance in parsecs is displayed on the vertical axis. The dashed line illustrates the observing boundary defined by the coronagraph's inner working angle (IWA). For systems within about 7 parsecs, an Earth-analog with a semi major axis of 1 au has the strongest likelihood of observing the full range of spectral features in this region. Conversely, for systems further than 7 parsecs the inner working angle will limit the detection of absorption features at longer wavelengths, for a similar analog. [Right] Simulated atmospheric absorption features for a Jupiter-like exoplanet at 5.2 AU. The same wavelength range is shown, with system distance on the vertical axis. The observing boundary defined by the outer working angle (OWA) is represented by the rounded cut off in the data. In contrast to the IWA, the outer working angle favors detections of planets (with larger semi major axes) in more distant planetary systems. A planet orbiting at the same distance as Jupiter would not become detectable with such a coronagraph unless the planetary system is at a minimum distance of approximately 2.5 parsecs.